\documentclass[a4paper,fleqn,usenatbib,useAMS]{mnras}

\usepackage[T1]{fontenc}
\usepackage{ae,aecompl}

\usepackage{graphicx}	
\usepackage{amsmath}	
\usepackage{amssymb}	
\usepackage{bm}

\usepackage[utf8]{inputenc}

\title[FDM halo substructure]{Substructure of fuzzy dark matter haloes}

\author[X. Du et al.]{
Xiaolong Du$^{1}$\thanks{E-mail: xiaolong@astro.physik.uni-goettingen.de.}, 
Christoph Behrens$^{1}$,
Jens C. Niemeyer$^{1}$
\\
$^{1}$Institut fuer Astrophysik, Georg-August Universit\"at Goettingen,
Friedrich-Hund-Platz 1, D-37075 Goettingen, Germany\\
}

\date{Accepted XXX. Received YYY; in original form ZZZ}

\pubyear{2016}

\begin{document}
\label{firstpage}
\pagerange{\pageref{firstpage}--\pageref{lastpage}}
\maketitle

\begin{abstract}
We derive the halo mass function (HMF) for fuzzy dark matter (FDM) by solving the excursion set problem explicitly with a mass-dependent barrier function, which has not been done before. We find that compared to the naive approach of the Sheth--Tormen HMF for FDM, our approach has a higher cutoff mass and the cutoff mass changes less strongly with redshifts.
Using merger trees constructed with a modified version of the Lacey \& Cole formalism that accounts for suppressed small scale power and the scale-dependent growth of
FDM haloes and the semi-analytic {\small GALACTICUS} code, we study the statistics of halo substructure
including the effects from dynamical friction and tidal stripping.
We find that if the dark matter is a mixture of cold dark matter (CDM) and FDM, there will be a suppression on the halo substructure on small scales which may be able to solve the Missing Satellites Problem faced by the pure CDM model. The suppression becomes stronger with increasing FDM fraction or decreasing FDM mass. Thus, it may be used to constrain the FDM model. 
\end{abstract}

\begin{keywords}
elementary particles -- cosmology: theory -- dark matter
\end{keywords}

\section{Introduction}\label{sec:Intro}

Searching for deviations from the predictions of the cold dark matter (CDM) model for hierarchical structure formation in the substructure of dark matter (DM) haloes is a promising way to constrain the properties of DM candidates. While CDM successfully explains the observed distribution of galaxies, clusters, and cosmic microwave background anisotropies on large scales described well by linear theory, possible discrepancies remain on fully collapsed, nonlinear scales. In particular, the Missing Satellites Problem \citep{Klypin:1999uc,Moore:1999nt}, the Cusp-Core Problem \citep[see][for a review]{deBlok:2009sp}, and the Too-Big-To-Fail Problem \citep{Boylan-Kolchin01072011} have received much attention. While they might indicate DM microphysics beyond pure CDM, recent results from cosmological simulations including CDM, baryons, and subresolution feedback models appear to be consistent with the observed halo substructure properties \citep{Wetzel:2016wro}. Among the observational strategies, gravitational lensing has been proposed as a particularly powerful tool to quantify the abundance and density profiles of both resolved \citep{Hezaveh2016} and unresolved \citep{Hezaveh2014} subhaloes. Further information on halo substructure can be obtained, for instance, from observations of more Milky Way satellites with galaxy surveys such as DES \citep{DES2016,Marsh2016} and from tidal stream observations \citep{Sanders2016}.

Various classes of modifications to the standard CDM model have been suggested for the suppression of small-scale structure. Warm dark matter \citep[WDM;][]{Bode:2000gq,Abazajian:2005xn} with non-negligible thermal motions and self-interacting DM \citep{Spergel:1999mh} with measurable non-gravitational interactions can be realized as variants of the weakly interacting massive particles (WIMPs) DM paradigm. On the other hand, light (sub-eV), coherently oscillating scalar fields also behave like CDM on large scales while showing new phenomenology on small scales, starting roughly at their virial velocity de Broglie wavelength \citep{Hu:2000ke,Sahni:1999qe,Woo:2008nn}. For particle masses around $10^{-22}$ eV, this length is of the order of several kpc, making this mass range sensitive to constraints from observations on galactic scales \citep{Lora:2011yc,Schive2014a,Lora:2014cya,Marsh2015b,Bozek2015,Schive2016,Sarkar2016,Calabrese2016}. Ultralight axions (ULAs) from string theory compactifications, produced by field misalignment before inflation, are prominent candidates for such particles \citep{Witten:1984dg,Svrcek:2006yi,Arvanitaki:2009fg,Marsh2015}. Here, we follow \citet{Hu:2000ke} and use the term `Fuzzy Dark matter' (FDM) that describes the general class of ultralight scalar fields with negligible self-interactions. As string theory can comfortably accommodate a very large number of ULAs with a broad range of masses \citep{Arvanitaki:2009fg}, mixed dark matter (MDM) models assuming a combination of heavier, CDM-like particles (for the purposes of galactic structure), and FDM give rise to equally plausible scenarios with suppressed small-scale structure \citep{Marsh2014}. This is the framework that we aim to explore further in this work.

At the linear level, coherently oscillating scalar fields give rise to a scale-dependent sound speed that produces a steep suppression of the transfer function below the corresponding Jeans length \citep{Khlopov01081985,Hu:2000ke}. \citet{Marsh2014} computed the Sheth--Tormen halo mass function \citep[HMF;][]{Sheth:1999mn} for MDM models by defining a mass-dependent critical density for collapse from a scale-dependent linear growth function. The resulting HMF drops off below a critical mass corresponding approximately to the respective Jeans mass, more or less steeply depending on the FDM fraction to total DM. In \citet{Bozek2015}, this HMF was used to constrain FDM with the observed UV luminosity function and from reionization history. 
In the context of halo substructure in the MDM model, the main limitation of this approach is the absence of non-linear effects on the evolution of subhaloes.

The nonlinear evolution of FDM was studied by \citet{Schive2016} and \citet{Sarkar2016} using collisionless $N$-body simulations with modified initial conditions, accounting for the Jeans-scale suppression of the transfer function but neglecting any later effects of quantum pressure on the halo structure itself. \citet{Schive2014a} performed a direct cosmological simulation of the Schr\"odinger--Poisson (SP) system of equations describing the non-relativistic dynamics of light scalar field DM with sufficient resolution to observe the formation of solitonic cores whose radius is governed by the de Broglie length evaluated at the halo virial velocity. This behaviour was further explored with the help of SP simulations of multiple merging solitons \citep{Schive2014b}. The solitonic cores were found to be embedded in NFW-like haloes and to obey a scaling relation with the total halo mass. \citet{Marsh2015b} and \citet{Calabrese2016} used these results for the cored density profiles to constrain FDM masses from observations of dwarf galaxies \citep{Walker2011}. 

Simulations of the SP equations with sufficient dynamical range to measure the HMF directly are computationally infeasible at present. Moreover, running a sufficient number of high-resolution $N$-body simulations with FDM initial conditions to obtain meaningful substructure statistics is equally impractical. As an alternative, semi-analytic modelling can be used to compute the subhalo properties by means of heuristic models for dynamical friction and tidal effects as previously demonstrated for the case of WDM by \citet{Pullen:2014gna}. This approach, which we will follow in this work, relies on a generalized extended Press--Schechter (EPS) formalism with a mass-dependent excursion set barrier for collapse to construct HMFs and merger trees implemented into the public semi-analytic galaxy formation code {\small GALACTICUS} \citep{Benson2012,Benson:2012su}. Its computational efficiency allows a wider exploration of parameter space than direct simulations. The cost is a higher level of simplification, especially regarding the assumed mass-dependent critical collapse overdensity and the nonlinear interaction of haloes during and after merging.

Using the semi-analytic approach outlined above, we focus in this paper on the subhalo population of synthetic FDM and MDM merger trees. In particular, we investigate how different phenomenologically motivated modifications of the subhalo modelling affect the population, and how the subhalo structure changes with FDM particle mass and the fraction of FDM in the MDM model.

This paper is structured as follows. In Section \ref{sec:mtrees}, we describe our method to generate synthetic trees obeying the FDM physics in terms of the formation and growth of FDM haloes. Section \ref{sec:hprofile} describes which modifications were made to account for the expected change in the density profile of FDM compared to CDM. In Section \ref{sec:satmodel}, we briefly describe the models for satellite haloes implemented in {\small GALACTICUS} and how these models were changed in order to include the expected phenomenology for FDM haloes, followed by our results in Section \ref{sec:results}. Finally, we summarize and discuss our findings in Section \ref{sec:conclusions}.

Throughout this paper, we assume the cosmological parameters from Planck 2015 \citep{Planck:2015xua}: $\Omega_{\mathrm{m}}=0.306$, $\Omega_{\Lambda}=0.694$, $h=0.6781$, $n_{\mathrm{s}}=0.9677$, and $\sigma_8=0.8149$.

\section{Construction of merger trees}\label{sec:mtrees}

\subsection{A new determination of the HMF for FDM}\label{sub_sec:HMF}

The Press--Schechter formalism \citep{Press:1973iz} has been widely used to compute the HMF. Despite its relative simplicity,
it has been found to be in reasonable agreement with $N$-body simulations of CDM \citep[see][for a review]{Monaco1998}.
Its consistent derivation goes back to \citet{Bond:1990iw}, who derived the EPS
HMF by solving the so-called excursion set problem. The basic idea is to relate the collapse of a DM halo to a
random walk across a barrier.

The fraction of mass in collapsed objects is assumed to be equal to the fraction of trajectories first upcrossing the barrier, $f(S)\mathrm{d}S$, where $f(S)$ is the probability distribution of a random walk first crossing a barrier with a height of $B(S)$ and $S$ is the mass variance of the smoothed density field. 
Then, the number density of collapsed haloes with mass $M$ at time $t$ per logarithmic mass bin, i.e. the HMF, can be written as
\begin{equation}
\frac{\mathrm{d} n(M,t)}{\mathrm{d}\ln M}=-\frac{\rho_{\mathrm{m}}}{M} f(S,t) S\,\frac{{\mathrm{d}}\ln S}{\mathrm{d}\ln M}.
\label{HMF_PS}
\end{equation} 
Here, $\rho_{\mathrm{m}}$ is the mass density.
The variance is given by
\begin{equation}
S(R)\equiv\sigma(R)^2=\frac{1}{2{\upi}^2}\int P(k) W(k,R)^2 k^2 \mathrm{d}k,
\label{variance}
\end{equation}
where $P(k)$ is power spectrum, $W$ is a window function, and $R$ is the filter scale. A commonly used window function
is the top-hap filter that can be written in the Fourier space as
\begin{equation}
W(k,R)=\frac{3}{(kR)^3}\left[\sin{(kR)}-kR\cos{(kR)}\right].
\label{W_KR}
\end{equation}
The halo mass $M$ of a collapsed object is then identified with the mass enclosed within the filter scale $R$:
\begin{equation}
M=\frac{4}{3}\upi\rho_{\mathrm{m}} R^3.
\end{equation}
We note that \citet{Bond:1990iw} use a sharp-$k$ filter to derive the EPS HMF. In
this case, there is no clear definition of mass corresponding to the filter scale. Thus, we use the top-hap filter in this work. Further discussions on different choices of window function for DM models with a suppressed power spectrum on small scales can be found in \citet{Schneider:2013ria} and \citet{Schneider:2014rda}.

Given the barrier function $B(S)$ that is often set to the critical overdensity for collapse, $\delta_{\mathrm{c}}$,
the first crossing distribution $f(S)$ can be solved from the following integral equation \protect\citep{Bond:1990iw,Benson:2012su}:
\begin{equation}
\int_0^S f(S')\mathrm{d} S'+\int_{-\infty}^{B(S)}P(\delta,S)\mathrm{d}\delta=1,
\label{f_int1}
\end{equation}
where
\begin{eqnarray}
&&\!\!\!\!\!\!\!\!\!\!\!\!\!\!\!\!\!\!\!\!\!\!\!\!P(\delta,S)=\frac{1}{\sqrt{2\upi S}} \exp\left(-\frac{\delta^2}{2 S}\right)\nonumber\\
&&\!\!\!\!\!\!\!\!\!\!\!\!-\int_0^S f(S')\frac{1}{\sqrt{2\upi (S-S')}} \exp\left[-\frac{(\delta-B(S'))^2}{2 (S-S')}\right]\mathrm{d}S'
\label{P}
\end{eqnarray}
is the probability for a trajectory to lie between $\delta$ and $\delta+\mathrm{d}\delta$ at variance $S$.
Substituting equation~(\ref{P}) into equation~({\ref{f_int1}}), and integrating over $\mathrm{d}\delta$, we obtain
\begin{equation}
\int_0^S f(S') {\rm erfc}\left[\frac{B(S)-B(S')}{\sqrt{2 (S-S')}}\right]\mathrm{d} S'={\rm erfc}\left[\frac{B(S)}{\sqrt{2 S}}\right],
\label{f_int2}
\end{equation}
where ${\rm erfc}$ is the complementary error function.
For a linear barrier function $B(S)=B_0+B_1 S$, equation~(\ref{f_int2}) can be solved analytically:
\begin{equation}
f(S)=\frac{B(0)}{\sqrt{2\upi S}}\exp\left[-\frac{B^2(S)}{2 S}\right]\frac{1}{S}.
\label{f_LB}
\end{equation}
For CDM at a specific redshift, $B(S)=\delta_{\mathrm{c}}^{\mathrm{cdm}}$ is constant, so $f(S)$ is given by equation~(\protect\ref{f_LB}).

For FDM, the barrier
function $B(S)=\delta_{\mathrm{c}}^{\mathrm{fdm}}(M(S))$ is nonlinear in $S$, i.e. the haloes with larger $S$ (smaller mass) are subject to a larger barrier. As in \citet{Marsh2014}, we define the mass dependent critical overdensity for collapse for
FDM as
\begin{equation}
\delta_{\mathrm{c}}^{\mathrm{fdm}}(M,z)=\frac{D_{\mathrm{cdm}}(z)}{D_{\mathrm{fdm}}(M,z)}\delta_{\mathrm{c}}^{\mathrm{cdm}}(z),
\label{delta_c_fdm}
\end{equation}
where $D(M,z)$ is the mass-dependent growth factor.

The relative amount of growth between CDM and FDM is calculated following \citet{Marsh2014}:
\begin{eqnarray}
\!\!\!\!\!\!\!\!\!G(k,z)\!\!\!\!\!\!&\equiv&\!\!\!\!\!\!\frac{D_{\mathrm{cdm}}(z)}{D_{\mathrm{fdm}}(M,z)}\nonumber\\
\!\!\!\!\!\!&=&\!\!\!\!\!\!\frac{\delta_{\mathrm{cdm}}(k,z)\delta_{\mathrm{cdm}}(k_0,z_{\mathrm{h}})}{\delta_{\mathrm{cdm}}(k,z_{\mathrm{h}})\delta_{\mathrm{cdm}}(k_0,z)}\frac{\delta_{\mathrm{fdm}}(k,z_{\mathrm{h}})\delta_{\mathrm{fdm}}(k_0,z)}{\delta_{\mathrm{fdm}}(k,z)\delta_{\mathrm{fdm}}(k_0,z_{\mathrm{h}})},
\label{G_k}
\end{eqnarray}
where $k_0=0.002h\rm{Mpc}^{-1}$ is a pivot scale, and $z_{\mathrm{h}}$ is chosen to be large enough so that at the relevant redshift the shape of CDM power spectrum has frozen in. We set $z_{\mathrm{h}}=300$ as in \citet{Marsh2014}.

Ignoring the possible time dependence in $G$, a fitting function for $G$ is given in \citet{Marsh:2016vgj} based on numerical results from AxionCAMB \citep[see][]{Marsh2014,Hlozek:2014lca}:
\begin{equation}
G(M)=h_{\mathrm{F}}(x)\exp\left(a_3 x^{-a_4}\right)+[1-h_{\mathrm{F}}(x)]\exp\left(a_5x^{-a_6}\right),
\label{G_k_fitting}
\end{equation}
where
\begin{eqnarray}
\!\!\!\!\!\!x\!\!\!\!&=&\!\!\!\!M/M_{\mathrm{J}},\\
\!\!\!\!\!\!h_{\mathrm{F}}(x)\!\!\!\!&=&\!\!\!\!\frac{1}{2}\{1-\tanh\left[M_{\mathrm{J}}(x-a_2)\right]\},\\
\!\!\!\!\!\!M_{\mathrm{J}}\!\!\!\!&=&\!\!\!\!10^8 a_1 \left(\frac{m_{\mathrm{a}}}{10^{-22}\rm{eV}}\right)^{-3/2}\left(\frac{\Omega_{\mathrm{m}} h^2}{0.14}\right)^{1/4}h^{-1}M_{\mathrm{\odot}}.
\end{eqnarray}
The best-fitting parameters are found to be $\{a_1,a_2,a_3,a_4,a_5,a_6\}=\{3.4,1.0,1.8,0.5,1.7,0.9\}$.

Hereafter, for FDM mass $m_{\mathrm{a}}=10^{-22}{\rm eV}$ and different FDM fractions, we will use the numerical transfer function at different redshifts for FDM kindly provided by D.J.E. Marsh to compute the critical collapse overdensity from equations (\ref{delta_c_fdm}) and (\ref{G_k}). For FDM masses other than $10^{-22}{\rm eV}$, we will use the fitting function for $G$, equation (\ref{G_k_fitting}), instead.

\begin{figure}
\includegraphics[width=\columnwidth]{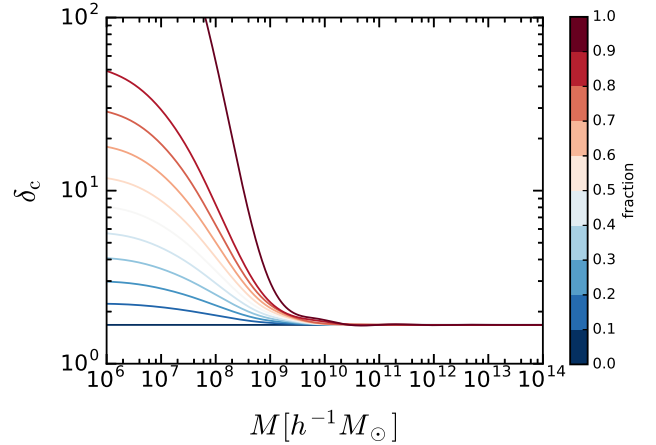}
\caption{Critical overdensity for collapse at $z=0$ with different FDM fractions $f$. The fractions range from 0 to 1 with a step size of $0.1$. The FDM mass is set to $10^{-22}\rm{eV}$.} 
\label{fig:delta_c_fractions}
\end{figure}

\begin{figure}
\includegraphics[width=\columnwidth]{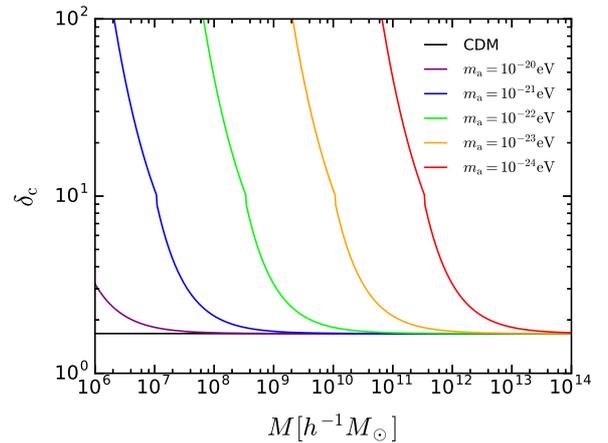}
\caption{Critical overdensity for collapse at $z=0$ with different FDM masses (based on the fitting formula from \protect\citealt{Marsh:2016vgj}) compared to standard CDM.} 
\label{fig:delta_c_masses}
\end{figure}

Figs \ref{fig:delta_c_fractions} and \ref{fig:delta_c_masses} show the critical overdensity for collapse at $z=0$ for different FDM fractions and different FDM masses, respectively.

Due to the non-linearity of the barrier function for
FDM, equation~(\ref{f_int2}) can only be solved numerically. We use a similar method as in \citet{Benson:2012su}.
Instead of a trapezoid rule for the integration, we implemented a mid-point rule that
is more stable for our case (see Appendix).

The original Press--Schechter formalism does not agree perfectly with $N$-body simulations of CDM. An improved model was proposed by \citet{Sheth:1999mn}:
\begin{equation}
f(S)= A\sqrt{\frac{1}{2\upi}}\sqrt{q}\nu\left[1+(\sqrt{q}\nu)^{-2p}\right]\exp\left(-\frac{q\nu^2}{2}\right)\frac{1}{S},
\label{ST}
\end{equation}
where $\nu\equiv \frac{B(S)}{\sqrt{S}}$, $A=0.3222$, $p=0.3$, and $q=0.707$. To be consistent with this fitting formula for $f(S)$, a remapping of the barrier was proposed by \citet{Sheth:1999su}:
\begin{equation}
B(S)\rightarrow B(S)\sqrt{q}\left[1+b\left(\frac{1}{q\,\nu^2}\right)^{c}\right],
\label{remap}
\end{equation}
where $b=0.5$ and $c=0.6$. Solving the integral equation~(\ref{f_int2}) with the remapped barrier function yields an HMF that is consistent with the Sheth--Tormen model.

In previous works \citep{Marsh2014,Bozek2015}, the HMF for FDM was computed using the Sheth--Tormen formula ( equation~\ref{ST}) 
with a redefined critical overdensity for collapse (equation~\ref{delta_c_fdm}).
However, the Sheth--Tormen formula was obtained for CDM whereas for a different barrier function, the solution for the first crossing distribution $f(S)$ should be different as discussed above. Strictly speaking, it is not self-consistent to substitute the barrier function for FDM directly into the solution of $f(S)$ for CDM. For this reason, we solve $f(S)$ from the integral equation~(\ref{f_int2}) using the numerical method described
above. To ensure that the HMF of FDM is consistent with that of CDM at higher masses, we keep the form of
the barrier remapping for CDM.

Note that the remapping, equation~(\ref{remap}), is derived and calibrated for CDM. For FDM haloes with mass $M$ much larger than the Jeans mass at matter-radiation equality $M_{\mathrm{J,eq}}$, this remapping is expected to work also for FDM since on these scales FDM behaves like CDM. But for FDM haloes with $M<M_{\mathrm{J,eq}}$, in which case FDM behaves very differently from CDM due to quantum pressure before the mass scale exceeds the Jeans mass at the corresponding time, this remapping may be inaccurate and needs to be recalibrated to simulations of FDM. As an approximation, in this work we use this remapping for all FDM haloes.

\subsection{Modification of the tree building algorithm}\label{sub_sec:build}

The merger trees are built using the \citet{Cole:2000ex} algorithm implemented in {\small GALACTICUS}.
First, a set of root haloes $\{M_i\}$ is drawn at redshift $z=0$. 
Given a root halo mass, the algorithm proceeds to construct a merger history by successively drawing branching events. The backwards evolution of a branch of the generated tree is considered complete if the relevant halo has a mass below the mass resolution limit which is a free parameter.

The merging rate is estimated from the conditional mass function computed with
the EPS formalism at two redshifts very close to each other assuming spherical collapse. Since the collapse of haloes is not perfectly
spherical, the merging rate calculated in this way does not accurately match the simulations. Therefore, we use the \citet{Parkinson:2007yh} modifications
to the merger rate
\footnote{A remapping of the barrier is used when calculating the HMF. But here as in \citet{Benson:2012su}, we use the \citet{Parkinson:2007yh} modifications to the merger rate instead of solving the excursion set problem with a remapped barrier.}.
The branching probability, i.e. the probability of a binary merger per
unit time, is then calculated by
\begin{equation}
P=\int_{M_{\mathrm{min}}}^{M/2}\frac{M}{M'}\frac{\mathrm{d}f}{\mathrm{d}t}\frac{\mathrm{d}S}{\mathrm{d}M'} G[\delta_{\mathrm{c}},
\sigma(M),\sigma(M')]\mathrm{d}M'.
\label{branching}
\end{equation}
Here, $M_{\mathrm{min}}$ is the lowest mass of resolved haloes and $G[\delta_{\mathrm{c}},\sigma(M),\sigma(M')]$ is an empirical modification obtained by \citet{Parkinson:2007yh}:
\begin{equation}
G[\delta_{\mathrm{c}},\sigma(M),\sigma(M')]=G_0\left(\frac{\sigma(M')}{\sigma(M)}\right)^{\gamma_1}\left(\frac{\delta_{\mathrm{c}}}{\sigma(M)}\right)^{\gamma_2},
\label{branching_mod}
\end{equation}
with $G_0=0.57$, $\gamma_1=0.38$ and $\gamma_2=-0.01$.
Note that this modification is calibrated to CDM $N$-body simulations, thus a recalibration may be needed for other
DM models. Nevertheless, we use the same form and parameters and leave the calibration for FDM to
future work when sufficiently large FDM simulations are available. As we show later, the HMF for FDM obtained from the merger trees fits the HMF derived above reasonably well even without a recalibration.

Additionally to binary mergers, the algorithm also accounts for smooth accretion. In the case of CDM, this corresponds to merging events in which the main halo accretes a small halo whose mass is below the mass resolution. For FDM, however, there is another source of smoothly accreted material, namely matter that is not locked up in a halo. The smooth accretion rate from this effect can be estimated by \citep{Benson:2012su} 
\begin{equation}
\left.\frac{\mathrm{d} R}{\mathrm{d}t}\right|_{\mathrm{smooth}}=\frac{\mathrm{d}f_{\mathrm{n}}}{\mathrm{d}t}G(\omega,\sigma(M),\sqrt{S_{\mathrm{max}}}).
\label{SAR}
\end{equation}
Here,
\begin{equation}
f_{\mathrm{n}}=1-\int_0^{S_{\mathrm{max}}} f(S) \mathrm{d}S.
\label{fn}
\end{equation}
\citet{Benson:2012su} find that if this effect is not included, the mass of haloes at higher redshifts will be overestimated for non-CDM models.

In addition to smooth accretion, we note that for FDM the probability to find any progenitors of a small halo with masses close to the HMF low-mass cutoff is very small. Therefore, we add an additional criterion for the termination of a tree branch: if the mass of a halo
is small and the probability of finding any of its progenitors is less than $P_{\mathrm{min}}$, the algorithm
will terminate the tree branch even if the node's mass is larger than the mass resolution.
Physically, such haloes correspond to those forming by collapsing density perturbations at that time as opposed to merging of
smaller haloes or smooth accretion.

Note that the branching probability criteria $P_{\mathrm{min}}$
must be chosen carefully. Like the HMF, the branching probability should drop off at small masses. Owing to numerical artefacts, the derived
branching probability becomes inaccurate when the progenitor mass approaches the cutoff, thus $P_{\mathrm{min}}$ must be chosen sufficiently large to avoid these artefacts. On the other hand, if $P_{\mathrm{min}}$ is too
large, the resulting HMF obtained from the merger trees will be affected, since the number of haloes close to the cutoff will be underestimated. Based on these two considerations, we choose different
$P_{\mathrm{min}}$ for different FDM fractions and masses. For example, for $f=1$ and $m_{\mathrm{a}}=10^{22}{\rm eV}$ we choose a value of $P_{\mathrm{min}}=2\times10^{-11}{\rm Gyr}^{-1}$.

\section{FDM halo profile}\label{sec:hprofile}

In a cosmological simulation based on the comoving SP equations, \citet{Schive2014a} found that FDM haloes contain distinct solitonic cores embedded in an NFW-like profile.
We can approximate the density profile of FDM haloes by
\begin{equation}
\rho_{\mathrm{FDM}}(r)=\Theta(r_{\epsilon}-r)\rho_{\mathrm{c}}(r)+\Theta(r-r_{\epsilon})
\rho_{\mathrm{NFW}}(r),
\label{FDM_profile}
\end{equation}
where $\Theta$ is the Heaviside step function, $r_{\epsilon}$ is the transition radius where the density profile changes from core to NFW-like behaviour, and the soliton density can be well described by \citep{Schive2014a,Schive2014b}
\begin{equation}
\rho_{\mathrm{c}}(r)=\frac{1.9 a^{-1}(m_{\mathrm{a}}/10^{-23}{\rm eV})^{-2}(a^{-1}r_{\mathrm{c}}/{\rm kpc})^{-4}}{[1+9.1\times10^{-2}(r/r_{\mathrm{c}})^2]^8}M_{\mathrm{\odot}}{\rm pc}^{-3}.
\label{sol_profile}
\end{equation}
Defining the core mass $M_{\mathrm{c}}$ as the mass enclosed within $r_{\mathrm{c}}$, \citet{Schive2014b} found a simple relation between the core mass and the halo mass,
\begin{equation}
M_{\mathrm{c}}\propto a^{-1/2} M_{\mathrm{h}}^{1/3}.
\label{schiveRe}
\end{equation}
It should be noted that by definition, $M_{\mathrm{c}}$ is at most $1/4$ of the halo mass.

In actual computations, we do not implement the full density profile as in equation (\ref{FDM_profile}), because
$r_{\epsilon}$ is still unknown \citep[see the discussion in][]{Marsh2015b}. Instead, we model FDM haloes with NFW density profiles with a modified concentration parameter and account for the presence of the core as part of the satellite model that will be described in Section \ref{sec:satmodel}.

\subsection{Concentration parameter}\label{sub_sec:modC}

The shape of the NFW profile is determined by the concentration parameter. haloes collapsing earlier when the average density of the Universe was larger are more concentrated.
As shown in Section \ref{sub_sec:HMF}, FDM haloes collapse at a higher critical overdensity than their CDM counterparts.
They form later than CDM haloes with the same mass, thus they are less concentrated. This is analogous to a similar effect seen in WDM; to account for it, we estimate the concentration parameter for FDM using a fitting formula derived for WDM \citep{Schneider:2011yu}:
\begin{equation}
\frac{c_{\rm{fdm}}}{c_{\rm{cdm}}}=\left(1+\gamma_1\frac{M_{1/2}}{M}\right)^{-\gamma_2},
\label{mod_c}
\end{equation}
where $\gamma_1=15$, $\gamma_2=0.3$, and $M_{1/2}$ is the half-mode mass defined according to the wavenumber at which the transfer function of FDM falls to one half of the transfer function for pure CDM. The concentration parameter for CDM in equation (\ref{mod_c})
is calculated according to \citet{Gao:2007gh}.
In the absence of large FDM simulations for calibrating the parameters
$\gamma_1$ and $\gamma_2$, we set them to the same value as for WDM \citep{Schneider:2011yu}. It is, however, a plausible assumption that FDM behaves more like collisional matter than CDM as a consequence of quantum pressure, so that stripping may be more effective.
In order to account for these uncertainties, we increased (decreased) the value of $\gamma_2$ by a factor of 2 (1/2) and adjusted $\gamma_1$ to match the concentration parameter at the half-mode mass. This variation gives rise to only an insignificantly small difference in the subhalo mass function (SHMF).
We also experimented with an alternative model for the concentration parameter based on \citet{Navarro:1995iw,Navarro:1996gj}, again finding only a small effect. The apparent insensitivity of the SHMF to changes of the concentration parameter justifies our use of equation~(\ref{mod_c}) with WDM parameters until more detailed FDM simulations become available.

\section{Satellite Model}\label{sec:satmodel}

The constructed merger trees yield information about hierarchical structure formation due to mergers. However, they do not include detailed modelling of the merging process, e.g. dynamical friction or tidal stripping. These processes require additional models to accurately follow the evolution of satellites from the time where they enter the virial radius of their hosts (which is the time at which they are considered to become satellites) until they dissolve within the host. 

{\small GALACTICUS} contains several different implementations of such models. The `simple' implementation only calculates the time at which the satellite will be completely merged with its host using a prescription from \citet{Jiang2008} calibrated to match CDM $N$-body simulations
\footnote{This is the default choice for the merging time-scales. However, {\small GALACTICUS} allows us to use several other implementations for the calculation of the merging time.}. 

The `orbiting' implementation presented in \citet{Pullen:2014gna} is a more sophisticated model. It includes tidal stripping, tidal heating, and integrates the satellites' orbits including a submodel for dynamical friction to derive the actual merging time. For details on the model, we refer the reader to \citet{Pullen:2014gna}; here, we briefly summarize the model ingredients. 

If an object becomes a satellite, it is assigned an orbit, drawing the orbital parameters from PDFs that have been derived from $N$-body simulations. We use {\small GALACTICUS}' default choice for these PDFs \citep[see][]{Benson2005}. The dynamics of the satellite in its host's gravitational potential is then evaluated, using a prescription for the acceleration ${\bm a}_{\mathrm{df}}$ by dynamical friction by \citet{Chandrasekhar1943}:
\begin{eqnarray}
{\bm a}_{\mathrm{df}} = -4\upi G^2 \, \ln \Lambda M_{\mathrm{sat}}\, \rho_{\mathrm{host}}(r_{\mathrm{sat}})\,\frac{{\bm V}_{\mathrm{sat}}}{V^3_{\mathrm{sat}}}\nonumber\\
\times \left[{\rm erf}(X)-\frac{2X}{\sqrt{\upi}} \exp(-X^2)\right],
\end{eqnarray}
where $M_{\mathrm{sat}}$ is the mass of the satellite, $\rho_{\mathrm{host}}(r_{\mathrm{sat}})$ is the density of the host halo at the distance of the satellite, ${\bm V}_{\mathrm{sat}}$ is the velocity of the satellite, $X = V_{\mathrm{sat}}/\sqrt{2}\sigma_{\mathrm{v}}$ with the velocity dispersion of the host $\sigma_{\mathrm{v}}$, and $\ln \Lambda$ is the Coulomb logarithm.

As a consequence of dynamical friction, the satellites' orbits shrink over time and they eventually dissolve. Satellites are considered merged if the distance is either lower than the sum of the half-mass radius of both satellite and host, or if the distance becomes less than $1$ per cent of the host's virial radius, or if the bound mass of the satellite is less than some fraction of its initial mass when becoming a satellite; this fraction is $1$ per cent by default. 

To account for tidal stripping, a tidal radius is defined so that all the mass outside the tidal radius is stripped from the satellite in one orbital period \citep{King1962}:
\begin{equation}
x_{\mathrm{t}}^3 = \frac{GM_{\mathrm{sat}}(< x_{\mathrm{t}})}{\omega^2-\mathrm{d}^2\Phi/\mathrm{d}r^2},
\end{equation}
with $M_{\mathrm{sat}}(< x_{\mathrm{t}})$ being the enclosed mass within the tidal radius, $\omega$ the angular velocity, and $\Phi$ the gravitational potential from the host. Finally, tidal heating is modelled using the results by \citet{Gnedin1999}.

\subsection{Modifications for FDM}\label{sub_sec:modFDM}

The existence of a core in the halo is a generic property of FDM models, changing the DM halo profile. While we expect these changes not to be significant at large radii, the close interaction of satellites and hosts with each other may potentially depend on the profile. In order to estimate the range of potentially observable effects, we use the satellite model to test several model assumptions, motivated in part by \citet{Schwabe2016} who ran detailed simulations of merging solitonic cores.

As a first assumption, we set the tidal stripping rate to zero once
the bound mass of the satellite approaches $4 M_{\mathrm{c}}$ (the total mass of an FDM halo when it is dominated by the solitonic core),
representing the idea that the compact core of the satellite is stable against tidal disruption. Secondly, we change the merging criteria. Host and satellite merge if the distance between them is smaller than the sum of their core radii, or if the satellite has less than $1$ per cent of its initial bound mass left. The criterion for the core radius is motivated by the observation that merging events happen rapidly once the cores touch \citep{Schwabe2016}. Physically, this is expected only to be relevant for small haloes that have relatively dominant cores while in more massive objects, one would expect the dynamics to be governed by the extended halo rather than the core. However, for these massive haloes the bound mass criterion will apply before the cores interact, yielding only mild changes in the SHMF at high masses (see section \ref{sec:results:smf}).

Additionally, the core mass is assigned along the merger history. First, we assign a core mass to all the haloes in a tree that have no progenitors according to the \citet{Schive2016} core mass--halo mass relation. Then the core mass may be changed by merger events.
Minor mergers, defined as mergers with core mass ratios higher than 7/3, do not affect the core mass of a halo. This is again motivated by \citet{Schwabe2016} who showed that at high core mass ratios, the core of the smaller halo is simply disrupted and does not add mass to the solitonic core. For major mergers (mergers with mass ratios smaller than 7/3), we set the core mass of the descendant halo $M_{\mathrm{c}}$ given the core masses of the progenitors, $M_{\mathrm{c,i}}$, to
\begin{equation}
M_{\mathrm{c}} = \beta \sum_i M_{\mathrm{c,i}},
\end{equation}
where $\beta$ parametrizes the mass loss of the cores in a merging event. A value of $\beta=1$ corresponds to the case where the core masses are simply summed up. We use a value of $\beta=0.7$ as found by \citet{Schwabe2016}.

As a last modification, we disable the tidal heating in the model. It is unclear how tidal heating would affect an FDM profile that has a core. In particular, the core itself should not expand at all. Therefore, we simply ignore tidal heating.

We note that we do not modify other parts of the satellite model. For instance, we use distributions of the orbital parameters of satellites calibrated from CDM simulations. Obtaining these distributions for FDM will require large cosmological simulations that are not available to date. In this work, we broadly explore the possible effects of FDM cores on these scales. A complete, self-consistent model is beyond the scope of this paper.

\section{Results}\label{sec:results}

The {\small GALACTICUS} code v0.9.4 (revision c49f04858120) is used with the modifications explained above (see Sections \ref{sec:mtrees}--\ref{sec:satmodel}). We compare the results for different FDM fractions $f=\Omega_{\mathrm{a}}/\Omega_{\mathrm{m}}$ and different FDM masses $m_{\mathrm{a}}$ with the standard CDM model.

\subsection{Halo Mass Function}\label{sec:results:hmf}

\begin{figure}
\includegraphics[width=\columnwidth]{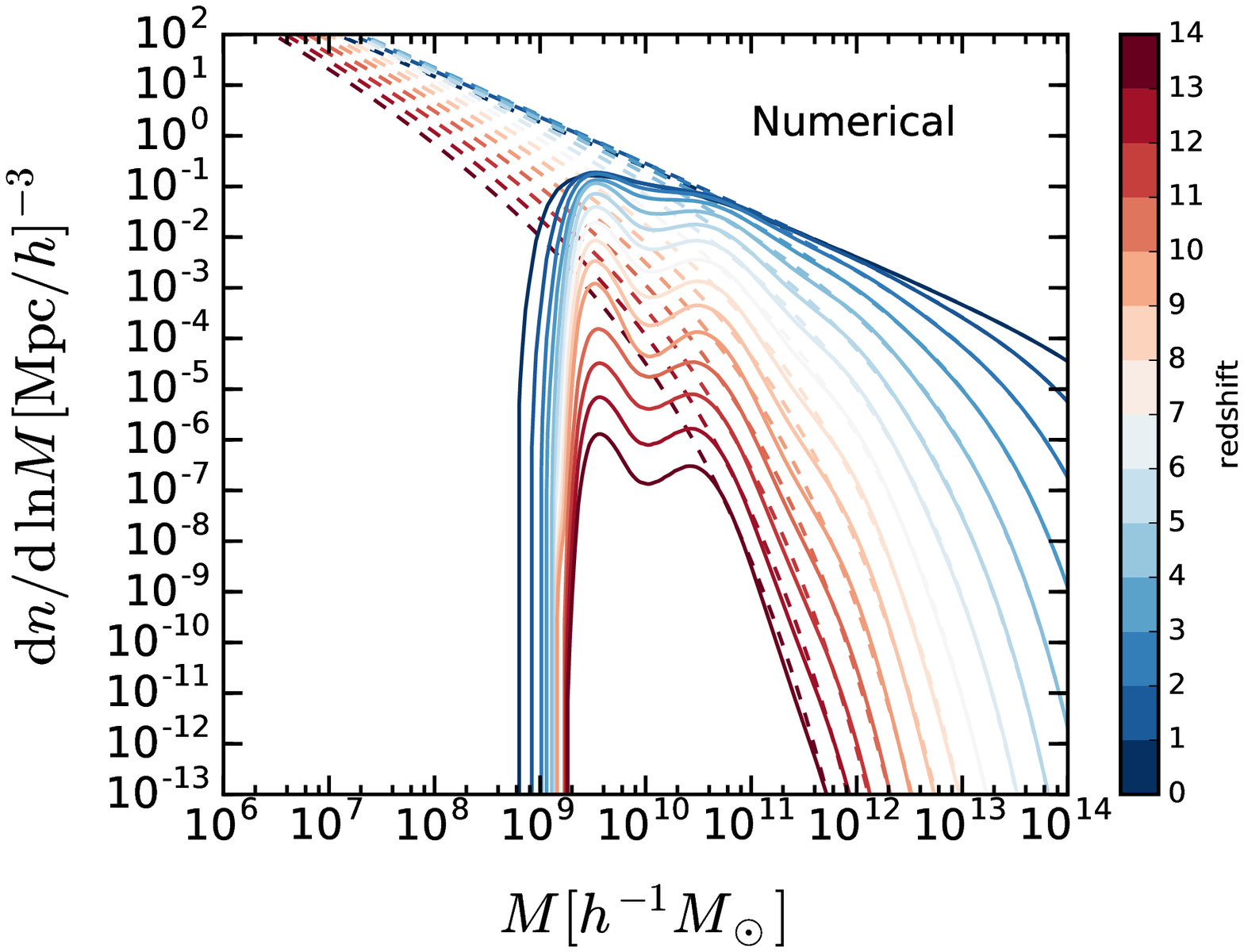}
\includegraphics[width=\columnwidth]{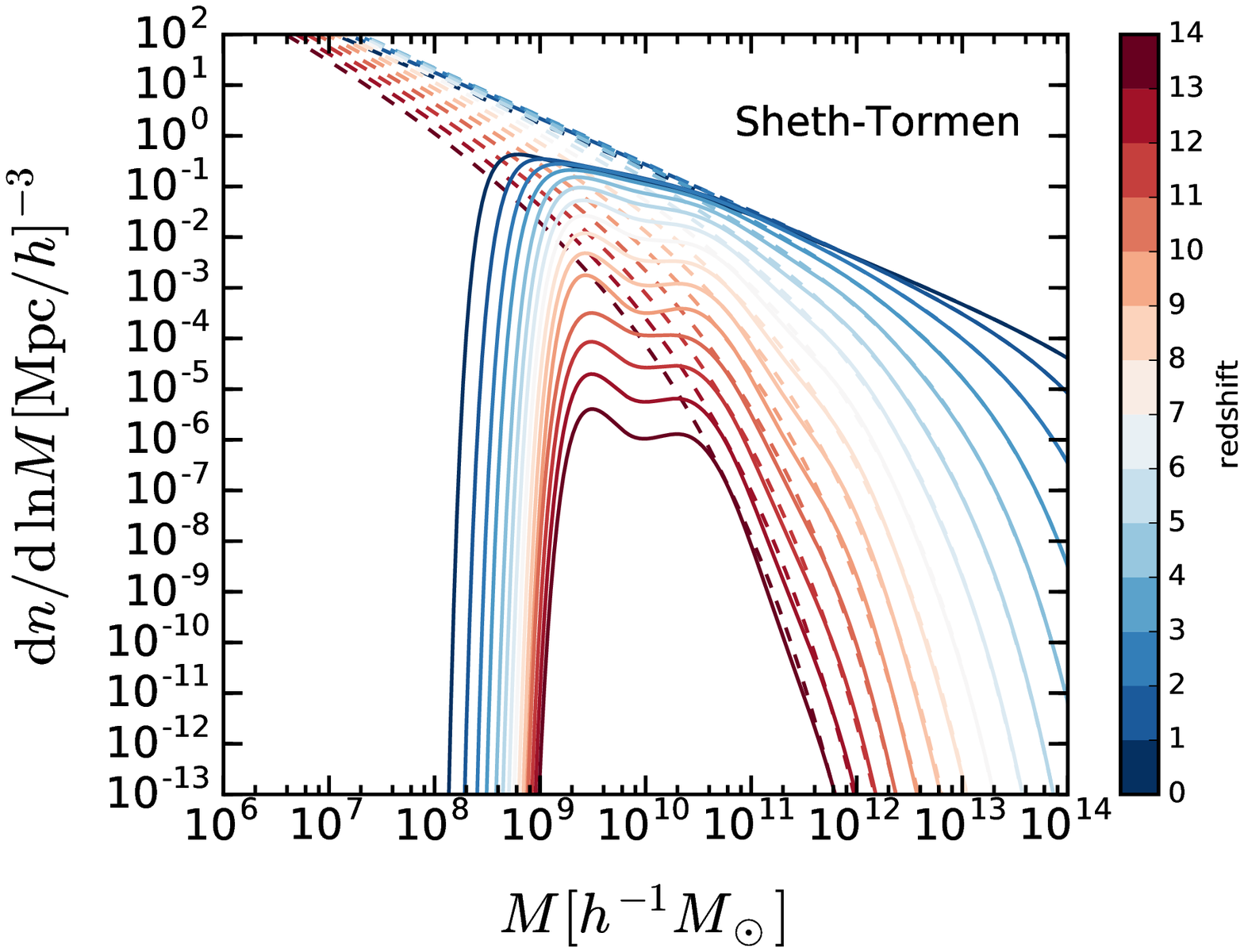}
\caption{HMF for FDM fraction $f = 1.0$ (solid lines) compared to standard CDM (dashed lines). Redshifts range from $z=14$ to $0$, obtained using our new calculation detailed above (upper panel) and the Sheth--Tormen formalism (lower panel). The FDM mass is set to $m_{\mathrm{a}}=10^{-22}{\rm eV}$.} 
\label{fig:HMF1p0}
\end{figure}

As discussed in Sec.~\ref{sub_sec:HMF}, we solve the excursion set problems for FDM with a redefined
critical overdensity for collapse for FDM. The HMF we obtain for pure FDM with $m_{\mathrm{a}}=10^{-22}{\rm eV}$ is shown in Fig.~\ref{fig:HMF1p0} (upper panel), with each line showing the HMF at a different redshift (see legend). 
We also show the FDM HMF as derived in previous investigations
\citep[e.g.][]{Bozek2015} by simply replacing the barrier
function in equation~(\ref{ST}) with the critical overdensity for collapse for FDM in the lower panel. 

As we can see from Fig.~\ref{fig:HMF1p0}, upper panel, the HMF of FDM we obtain numerically shows the characteristic cutoff below a redshift-dependent minimum mass owing to quantum pressure. Compared with the Sheth--Tormen formalism (lower panel), we obtain a higher
cutoff.
At $z=0$, the cutoff mass is about $6\times10^8 h^{-1} M_{\mathrm{\odot}}$, roughly four times the value obtained from the Sheth--Tormen formalism. Additionally, the cutoff mass of the HMF changes less strongly with redshift. At $z=14$, the cutoff mass is $2\times10^9 h^{-1} M_{\mathrm{\odot}}$, only two times of the value obtained from the Sheth--Tormen formalism.

\begin{figure}
\includegraphics[width=\columnwidth]{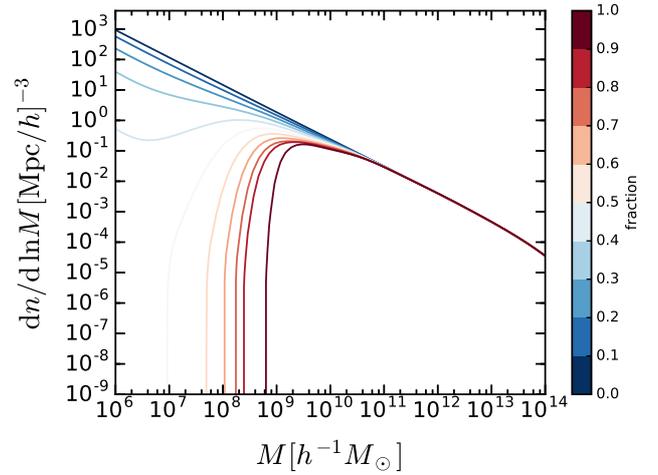}
\caption{HMF at $z=0$ with different FDM fractions $f$. The fractions range from 0 to 1 with a step size of $0.1$. The FDM mass is set to $m_{\mathrm{a}}=10^{-22}{\rm eV}$.} 
\label{fig:HMF_fractions}
\end{figure}

Next, we consider the case that DM is composed of a mixture of CDM and FDM (where the CDM component may simply be another ULA species with higher mass). We fix the total matter density of DM and change the fraction of
FDM, $f=\Omega_{\mathrm{a}}/\Omega_{\mathrm{m}}$, from $0$ (pure CDM) to $1$ (pure FDM). The HMFs at $z=0$ for different fractions is shown in Fig.~\ref{fig:HMF_fractions}. We find that on large scales, the HMFs for MDM models are consistent with CDM as expected. However, on small scales
the HMF is suppressed. With increasing $f$, the suppression becomes more and more significant. Around equipartition between CDM and FDM ($f\sim0.5$), we recover a sharp cutoff as in the pure FDM case, albeit at lower cutoff masses ($\sim10^7 h^{-1} M_{\mathrm{\odot}}$). For higher $f$, the cutoff mass increases. 
All of these results are, mostly by construction, qualitatively consistent with \citet{Marsh2014}.

\begin{figure}
\includegraphics[width=\columnwidth]{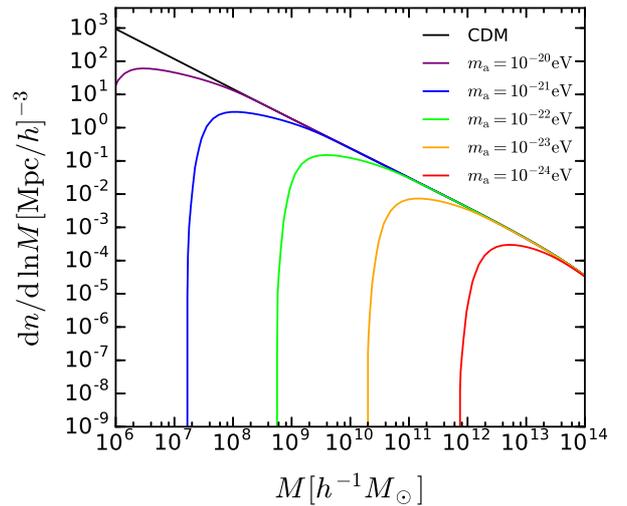}
\caption{HMF at $z=0$ with different FDM masses compared to standard CDM.} 
\label{fig:HMF_masses}
\end{figure}

Finally, we also consider different particle masses for FDM,
using fitting functions for the transfer function, equation (8)
in \citet{Hu:2000ke}, and $G$,  equation (\ref{G_k_fitting}) \citep{Marsh:2016vgj}. 
 The results are shown in Fig. \ref{fig:HMF_masses}. We find the cutoff in the HMF changing with FDM mass as expected. The smaller the FDM mass, the larger the cutoff; e.g. for $m_{\mathrm{a}}=10^{-24}{\rm eV}$, the cutoff mass is about $10^{12}M_{\mathrm{\odot}}$, a possibility clearly ruled out by the existence of Milky Way-sized haloes.

\subsection{Validating Merger Trees}\label{sec:results:validtrees}

\begin{figure}
\includegraphics[width=\columnwidth]{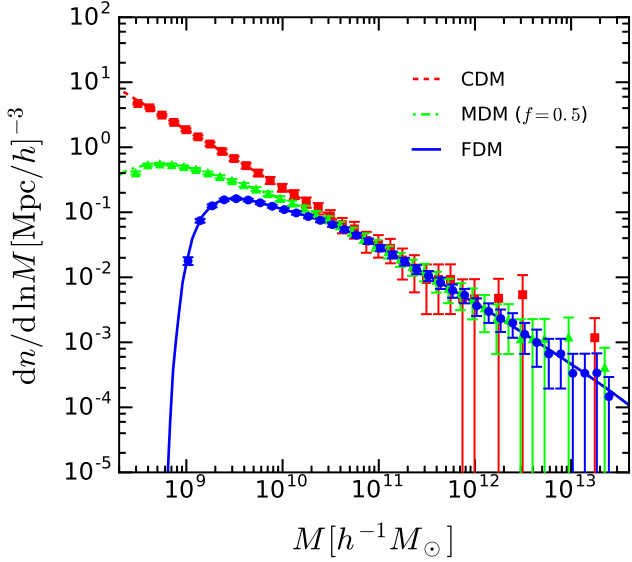}
\caption{HMF from merger trees at $z=0$ for pure FDM and MDM compared to standard CDM. The FDM mass is set to $m_{\mathrm{a}}=10^{-22}{\rm eV}$. Symbols indicate the data from the merger trees, the lines indicate the directly calculated HMF.} 
\label{fig:HMF_MergerTree0}
\end{figure}

To check the consistency of the merger tree algorithm, we run merger trees with $1000$ trees per decade in mass for haloes with masses in the range
$[4\times10^{8},4\times10^{13}]\,M_{\mathrm{\odot}}$ at $z=0$. The mass resolution is set to $2\times10^8 M_{\mathrm{\odot}}$.
We compare the HMF obtained by counting the haloes in our merger trees with
the one derived from solving the excursion set problems at different redshifts. Fig.~\ref{fig:HMF_MergerTree0} shows the HMF at $z=0$ for pure FDM, $f=0.5$, and pure CDM, with the symbols showing the data while the lines show the expected HMF from directly solving the excursion set problem as discussed in the previous sections.
In order to show the points and error bars more clearly and avoid overlapping, the centre of bins is chosen slightly differently when counting the haloes in merger trees. We can see that the HMFs obtained from merger trees can reproduce the excursion set problem solutions.
The large errors at higher masses are due to relatively few large haloes.

\begin{figure}
\includegraphics[width=\columnwidth]{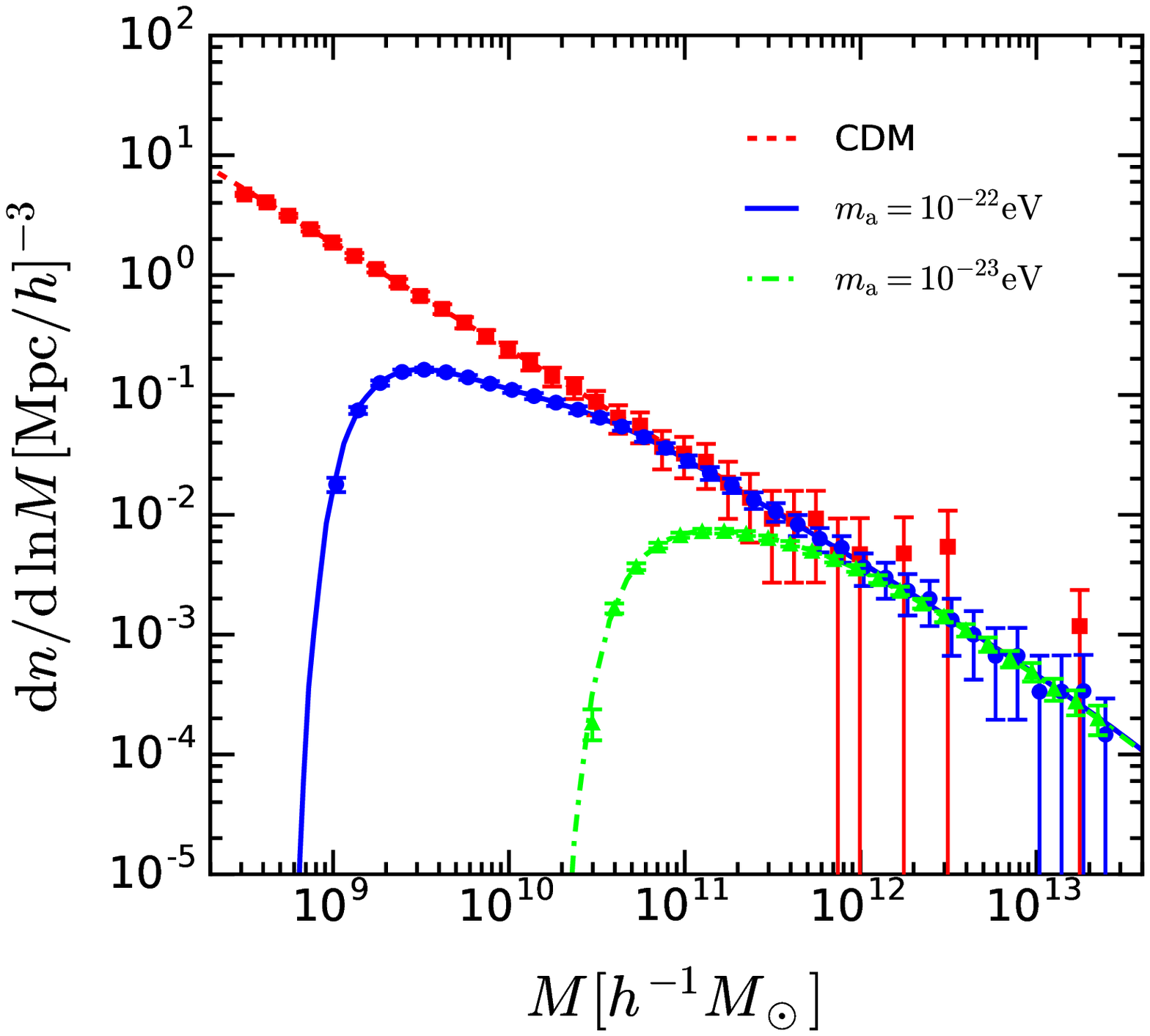}
\caption{HMF from merger trees at $z=0$ for pure FDM with different masses compared to standard CDM.} 
\label{fig:HMF_MergerTree_Masses0}
\end{figure}

Similarly, Fig.~\ref{fig:HMF_MergerTree_Masses0} shows the HMF obtained from the merger trees compared to the direct solution for pure CDM and two different FDM masses. The agreement with the expectations is good.

\begin{figure}
\includegraphics[width=\columnwidth]{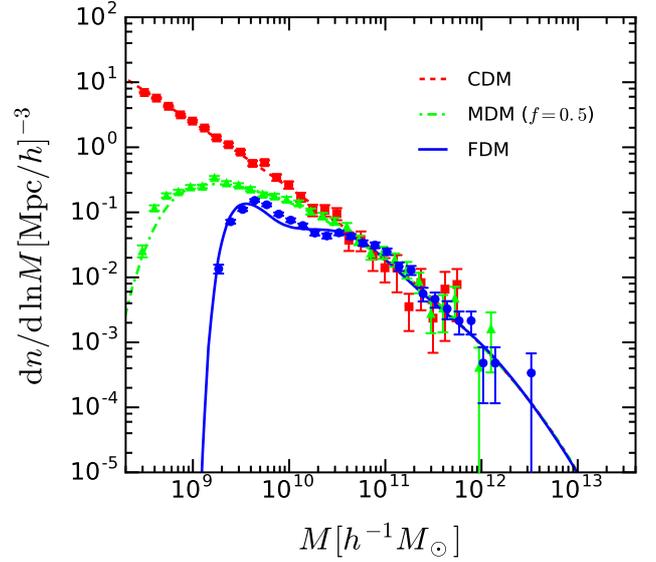}
\caption{HMF from merger trees at $z=3$ for pure FDM and MDM compared to standard CDM. The FDM mass is set to $m_{\mathrm{a}}=10^{-22}{\rm eV}$.} 
\label{fig:HMF_MergerTree3}
\end{figure}

\begin{figure}
\includegraphics[width=\columnwidth]{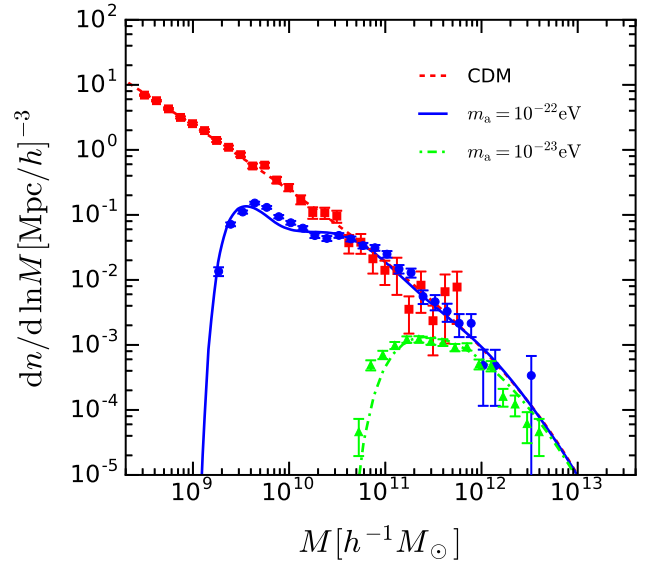}
\caption{HMF from merger trees at $z=3$ for pure FDM with different masses compared to standard CDM.} 
\label{fig:HMF_MergerTree_Masses3}
\end{figure}

Fig.~\ref{fig:HMF_MergerTree3} (Fig.~\ref{fig:HMF_MergerTree_Masses3}) again shows the HMF from the merger trees for different fractions $f$ (FDM masses), but for a different redshift of $z=3$. We find an acceptable agreement, despite small deviations at intermediate halo masses. At even larger redshifts, the HMFs derived from the merger trees deviate more from the excursion set solutions. This is caused by small deviations in the branching modelling that accumulate with redshift, since the tree generation algorithm works `top to bottom', i.e. it starts at $z=0$ and generates progenitors. This deviation suggests that we may need to recalibrate the \citet{Parkinson:2007yh} modifications to the merger rate for FDM to get more accurate merger histories. Since there exist no sufficiently large cosmological simulations for FDM at present, we leave this to future work.
We will show that this deviation at higher redshifts does not have a significant effect on the substructures of haloes at lower redshifts.

\subsection{Halo Substructure}\label{sec:results:smf}

In the last subsection, we have presented the HMFs for FDM and MDM. In this subsection, we will use the satellite models described in Section \ref{sec:satmodel} to explore the SHMF.

We ran merger trees with $1000$ trees per decade in mass for Milky Way-sized parent haloes, $1\times10^{12}<M_{\mathrm{host}}<3\times10^{12} M_{\mathrm{\odot}}$.
We chose a mass resolution of $M_{\mathrm{res}}=5\times10^7 M_{\mathrm{\odot}}$ and compared the results from different satellite models, the `simple' implementation and the `orbiting' implementation. For the pure FDM case, we use the orbiting model with the modifications outlined above; for CDM, we use the unmodified version. In the orbiting implementation, we used the same parameters as in \citet{Pullen:2014gna}: the Coulomb logarithm
$\ln\,\Lambda=2.0$ and the tidal stripping mass loss rate parameter $\alpha=2.5$. The model for tidal heating in the orbiting implementation was switched off since it is physically unclear how this is to be treated given that FDM haloes are expected to have compact cores.

\begin{figure}
\includegraphics[width=\columnwidth]{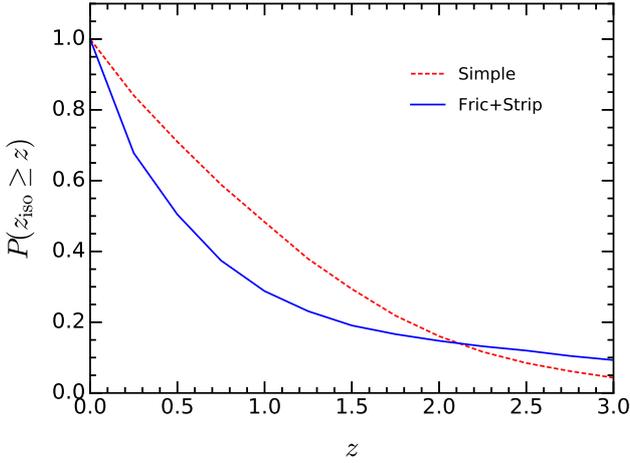}
\caption{Cumulative distribution of the redshifts at which $z=0$ subhaloes became subhaloes for pure FDM ($f=1$) with $m_{\mathrm{a}}=10^{-22}{\rm eV}$.} 
\label{fig:LastIsolated}
\end{figure}

As discussed in last subsection, the HMF obtained from merger trees may deviate from the one from solving the excursion set problems at redshift $z>3$.
It indicates that the merger tree structure at high redshifts suffers from the uncertainty caused by an inaccurate calculation of merging rate. In order to check how significantly this affects
the SHMF, we use the redshift at which subhaloes were last isolated as a diagnostic. As long as most of the subhaloes with $z_{\mathrm{iso}}>3$ have been completely merged in their host at $z=0$ (and therefore are not present in the substructure of their hosts), the high-redshift HMF does not affect the SHMF at low redshift. 
In Fig.~\ref{fig:LastIsolated}, we show the cumulative fraction of subhaloes with last isolated redshift $z_{\mathrm{iso}}\ge z$ for the case with $f=1$ and $m_{\mathrm{a}}=10^{-22}{\rm eV}$. As can be seen, less than $10$ per cent of subhaloes originate from isolated haloes at redshifts larger than $3$. Thus, the uncertainty at higher redshifts can only have a very small effects on the SHMF at $z=0$.

\begin{figure}
\includegraphics[width=\columnwidth]{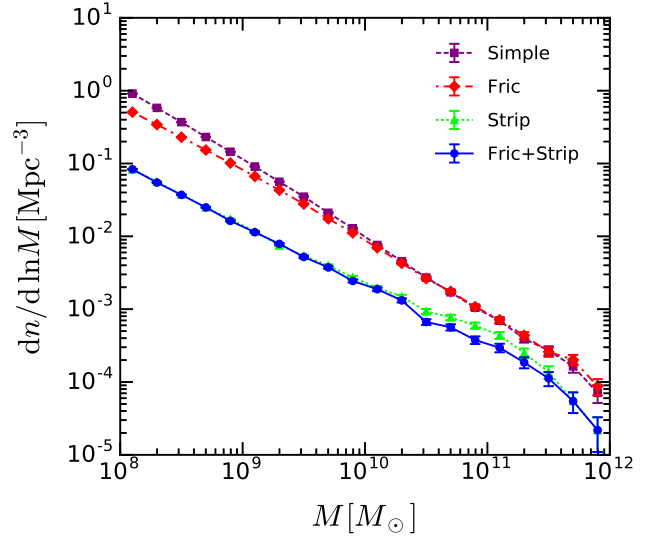}
\caption{SHMF from merger trees for standard CDM. `Fric' refers to dynamic friction and `Strip' refers to tidal stripping.} 
\label{fig:SHM_MergerTrees_0p0}
\end{figure}

To show the effects of dynamical friction and tidal stripping in the orbiting model separately, we show the SMHF with each effect switched on and compare them with the simple model. We note that in the following, we use the concentration parameter as defined for CDM (see above). We will analyse the influence of the concentration parameter on the stripping efficiency and therefore on the SMHF below. Fig.~\ref{fig:SHM_MergerTrees_0p0} shows the results for CDM, similar to figs 2 and 3
in \citet{Pullen:2014gna}. Due to changes between version v0.9.3 and v0.9.4 in {\small GALACTICUS}, 
there exist some small differences but the results are still comparable. 
We can see that if only dynamical friction is active, the SHMF is broadly consistent with the simple model. The tidal stripping effect reduces the amplitude of the SHMF. This is because the mass bound to the subhaloes is gradually stripped by tidal forces, shifting the SHMF to the left which in turn makes the subhaloes more vulnerable to further stripping. This is consistent with the results of \citet{Pullen:2014gna}.

\begin{figure}
\includegraphics[width=\columnwidth]{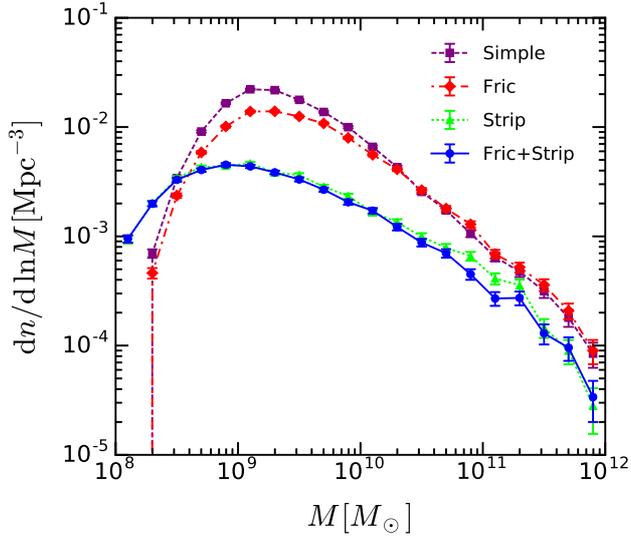}
\caption{SHMF from merger trees for MDM ($f=0.5$). The FDM mass is set to $m_{\mathrm{a}}=10^{-22}{\rm eV}$.} 
\label{fig:SHM_MergerTrees_0p5}
\end{figure}

\begin{figure}
\includegraphics[width=\columnwidth]{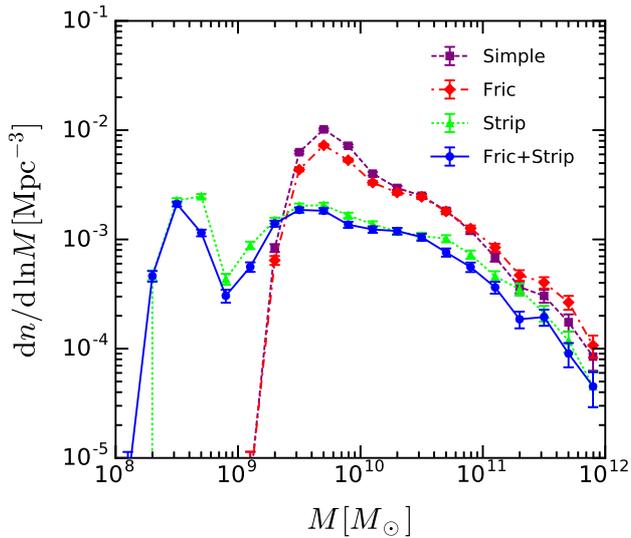}
\caption{SHMF from merger trees for pure FDM ($f=1$). The FDM mass is set to $m_{\mathrm{a}}=10^{-22}{\rm eV}$.} 
\label{fig:SHM_MergerTrees_1p0}
\end{figure}

Figs \ref{fig:SHM_MergerTrees_0p5} and \ref{fig:SHM_MergerTrees_1p0} show the SHMF for MDM with $f=0.5$ and pure FDM, respectively.
The FDM mass is taken to be $10^{-22}{\rm eV}$.
For the MDM case, we use the original orbiting implementation, i.e. we do not include the effects arising from axionic cores. As can be seen, if some fraction of DM is FDM,
the SHMF is suppressed at lower masses.

In the simple satellite model and the orbiting model
with only dynamical friction, subhaloes do not lose mass until they dissolve within the host halo. In these models, we therefore observe a sharp
cutoff in the SHMF at the HMF mass cutoff. In the more realistic orbiting model including tidal stripping, the subhaloes will gradually lose their
mass due to tidal stripping. This effect can be identified in Figs \ref{fig:SHM_MergerTrees_0p5} and \ref{fig:SHM_MergerTrees_1p0}; the SMHF obtained from the orbiting model including
tidal stripping is shifted to the lower mass end. In particular, for the pure FDM case, we can see a bump that peaks at $3-4\times10^8 M_{\mathrm{\odot}}$ if we include tidal stripping in the orbiting model. This is expected as a consequence of our modification to stop tidal stripping at the solitonic core (cf. Section \ref{sec:satmodel}). These naked cores orbit their hosts until they satisfy one of the merging criteria, yielding a bump in the SMHF around
$4M_{\mathrm{c}}$ (the factor $4$ follows from the definition of the core mass). At present, this feature is a direct consequence of our model assumptions. Large simulations resolving subhalo structure and dynamics are required to determine whether the feature in the SMHF is physical. 

\begin{figure}
\includegraphics[width=\columnwidth]{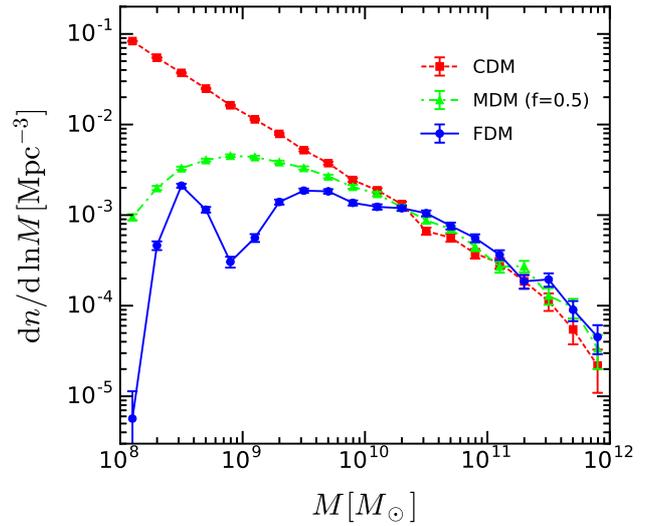}
\caption{SHMF for FDM and MDM ($f=0.5$) using the `orbiting' implementation with dynamical friction and tidal stripping compared with the standard CDM. The FDM mass is set to $m_{\mathrm{a}}=10^{-22}{\rm eV}$.} 
\label{fig:SHM_MergerTrees_Com}
\end{figure}

Fig.~\ref{fig:SHM_MergerTrees_Com} shows a comparison of the SHMFs for FDM, MDM ($f=0.5$), and CDM using the orbiting implementation with dynamical fraction and tidal stripping switched on. It can be seen that the SHMFs for FDM and MDM are suppressed at smaller masses compared to CDM, while at larger masses the three models are coincident.

\begin{figure}
\includegraphics[width=\columnwidth]{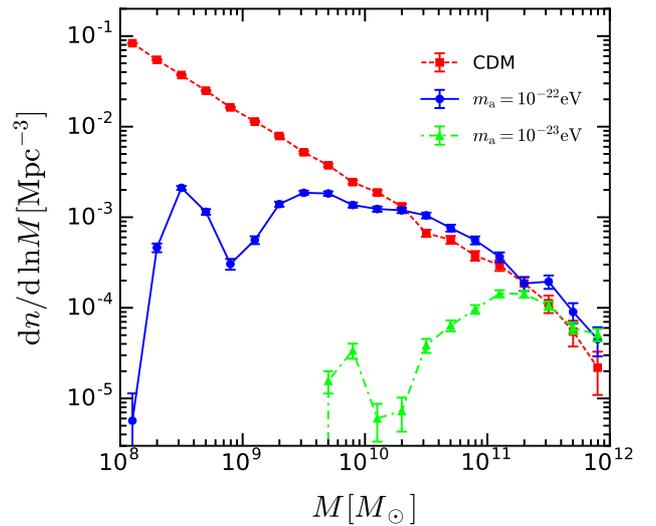}
\caption{SHMF for FDM with different masses compared to standard CDM.} 
\label{fig:SHM_MergerTrees_Masses_Com}
\end{figure}

We also compare the SHMF for FDM with different masses in Fig. \ref{fig:SHM_MergerTrees_Masses_Com}. For the case of $m_{\mathrm{a}}=10^{-23}{\rm eV}$, we ran the merger
trees with $4000$ trees per decade in mass in order to reduce the statistical errors close to the cutoff.
With decreasing FDM mass, there are fewer lower mass subhaloes as expected from the increased mass cutoff in the HMF for lower axion masses.

\begin{figure}
\includegraphics[width=\columnwidth]{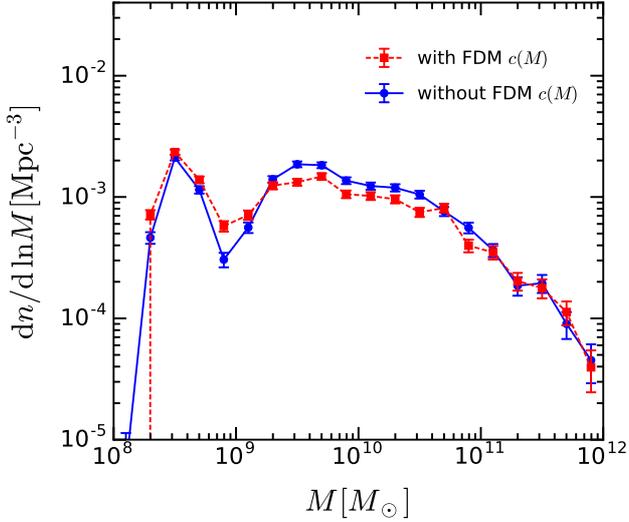}
\caption{SHMF for FDM with and without modification to concentration parameter. The FDM mass is set to $m_{\mathrm{a}}=10^{-22}{\rm eV}$.} 
\label{fig:SHM_MergerTrees_modc}
\end{figure}

In Fig. \ref{fig:SHM_MergerTrees_modc}, we show the results for SMHF for two different choices of the concentration parameter as described in Section \ref{sub_sec:modC}. The differences are marginal. However, for the modified concentration parameter, the lower concentration parameter make haloes more susceptible to tidal stripping. Consequently, the SMHF in the mass range $2\times10^9\sim10^{11}M_{\mathrm{\odot}}$ is slightly lower.

\section{Conclusions}\label{sec:conclusions}

In this work, we describe the implementation of FDM into the semi-analytic code for galaxy formation {\small GALACTICUS} in order to study the substructure of FDM haloes. Using the FDM transfer function with a sharp small-scale cutoff caused by quantum pressure \citep{Hu:2000ke,Marsh2014,Hlozek:2014lca,Marsh:2016vgj} and a modified barrier function accounting for the mass dependent growth of FDM \citep{Marsh2014}, we apply the EPS formalism to calculate the HMF for FDM.

Our HMFs show minor differences to those derived with the Sheth--Tormen formalism with a redefined critical overdensity for collapse \citep{Benson:2012su,Marsh2014} by having a larger cutoff mass that changes less strongly with redshift (cf. Fig. \ref{fig:HMF1p0}).

Using the \citet{Cole:2000ex} algorithm implemented in {\small GALACTICUS}, we build synthetic merger trees for FDM and validate them by comparing the HMF
with the solution of the excursion set approach at different redshifts. We find that the HMFs  match reasonably well at redshift $z<3$ (see Figs. \ref{fig:HMF_MergerTree0}--\ref{fig:HMF_MergerTree_Masses3}), while deviations exist at higher redshifts. We demonstrate, however, that the vast majority of subhaloes accreted by the host at redshift $z>3$ are already completely merged with the host at $z=0$ and thus do not have a significant effect on the SHMF at the present time.

To study the nonlinear evolution of FDM subhaloes, we modify an existing model for satellite haloes (the `orbiting' implementation) in {\small GALACTICUS} \citep{Pullen:2014gna} which includes the nonlinear effects of dynamical friction and tidal stripping (we ignore tidal heating in the absence of calibrations with FDM simulations). In order to account for possible effects from compact solitonic cores in FDM haloes, we set the tidal stripping to $0$ when a satellite has only its core left
and change the merging criteria of subhaloes.
We find that if FDM composes a significant fraction of the total DM density, the SHMF is suppressed on small scales.
The larger the fraction or the smaller the mass, the stronger is the suppression (see Figs. \ref{fig:SHM_MergerTrees_Com} and \ref{fig:SHM_MergerTrees_Masses_Com}). For pure FDM, the SHMF exhibits a maximum around $4 M_{\mathrm{c}}$ (see Fig. \ref{fig:SHM_MergerTrees_Masses_Com}), where $M_{\mathrm{c}}$ is the mass of the solitonic core for subhaloes close to the mass cutoff, reflecting the fact that pure cores are assumed to be stable against tidal stripping.
We also consider the impact of modifying the concentration parameter for FDM haloes on to the SMHF and find only weak differences.

The parameters for dynamical friction and tidal stripping used in our work were obtained from CDM $N$-body simulations. A recalibration of these parameters will be possible as soon as sufficiently large FDM simulations become available.
In addition, we only considered gravitational interactions between the subhaloes and their host. Simulations of FDM \citep{Schive2014a,Schive2014b} show that FDM haloes have a granular structure caused by wave interference that might make the dynamics of FDM subhaloes behave differently from the collisionless case. In particular, stripping may be more efficient than expected from purely gravitational tidal effects\footnote{We thank the referee for pointing this out.}.

In this work, we have only explored pure DM physics. Using the generated merger trees in conjunction with a semi-analytic model of galaxy formation that includes baryonic physics, we will be able to constrain FDM models from observations like the high-redshift UV luminosity function and reionization history. Recently, \citet{Cen:2016htm} showed that a joint analysis of the Ly$\alpha$ forest, the mean free path of ionizing photons, and the galaxy luminosity function puts a stringent constraint on the upper bound of the HMF cutoff mass, which can possibly be used to constrain the mass of FDM or the fraction of FDM in MDM models.

\section*{Acknowledgements}
We thank D.J.E. Marsh, B. Schwabe, and J. Veltmaat for helpful discussions and D.J.E. Marsh for providing the FDM transfer function. We would also like to thank A. Benson for extensive discussion and support
and the referee, T. Chiueh, for constructive comments which helped to improve this paper.
XD acknowledges the China Scholarship Council (CSC) for financial support.

\appendix

\section{Numerical method for solving the integral equation}\label{apdix:numeric_int}
In \citet{Benson:2012su}, the integral equation~(\ref{f_int2}) is solved by discretizing the integral using the
trapezoid rule as
\begin{eqnarray}
&&\!\!\!\!\!\!\!\!\!\!\!\!\!\!\!\!\!\!\!\!\!\!\!\!\int_0^{S_j} f(S')K(S_j,S')\mathrm{d}S'\nonumber\\
&&\!\!\!\!\!\!\!\!\!\!\!=\sum_{i=0}^{j-1} \frac{f(S_i)K(S_j,S_i)+f(S_{i+1})K(S_j,S_{i+1})}{2}\Delta S_i.
\label{trapezoid_rule}
\end{eqnarray}
Here, $K(S_j,S')$ is the kernel of the integral equation. To increase precision for our specific problem, we instead implement the mid-point rule:
\begin{equation}
\int_0^{S_j} f(S')K(S_j,S')\mathrm{d}S'=\sum_{i=0}^{j-1}f(S_{i+1/2})K(S_j,S_{i+1/2})\Delta S_i.
\label{midpoint_rule}
\end{equation}
The first crossing distribution $f(S)$ at $S_{j-1/2}$ is given by
\begin{eqnarray}
&&\!\!\!\!\!\!\!\!\!\!\!\!\!\!\!\!\!\!\!\!\!\!\!\!\!\!\!\!\!f(S_{j-1/2})=\frac{1}{K(S_j,S_{j-1/2})\Delta S_{j-1/2}}\nonumber\\
&&\!\!\!\!\!\!\!\!\!\!\!\!\!\!\!\!\!\!\!\!\!\!\!\!\left({\rm erfc}\left[\frac{B(S_j)}{\sqrt{2 S_j}}\right]-\sum_{i=0}^{j-2}f(S_{i+1/2})K(S_j,S_{i+1/2})\Delta S_i\right).
\label{midpoint_sol}
\end{eqnarray}

\begin{figure}
\includegraphics[width=\columnwidth]{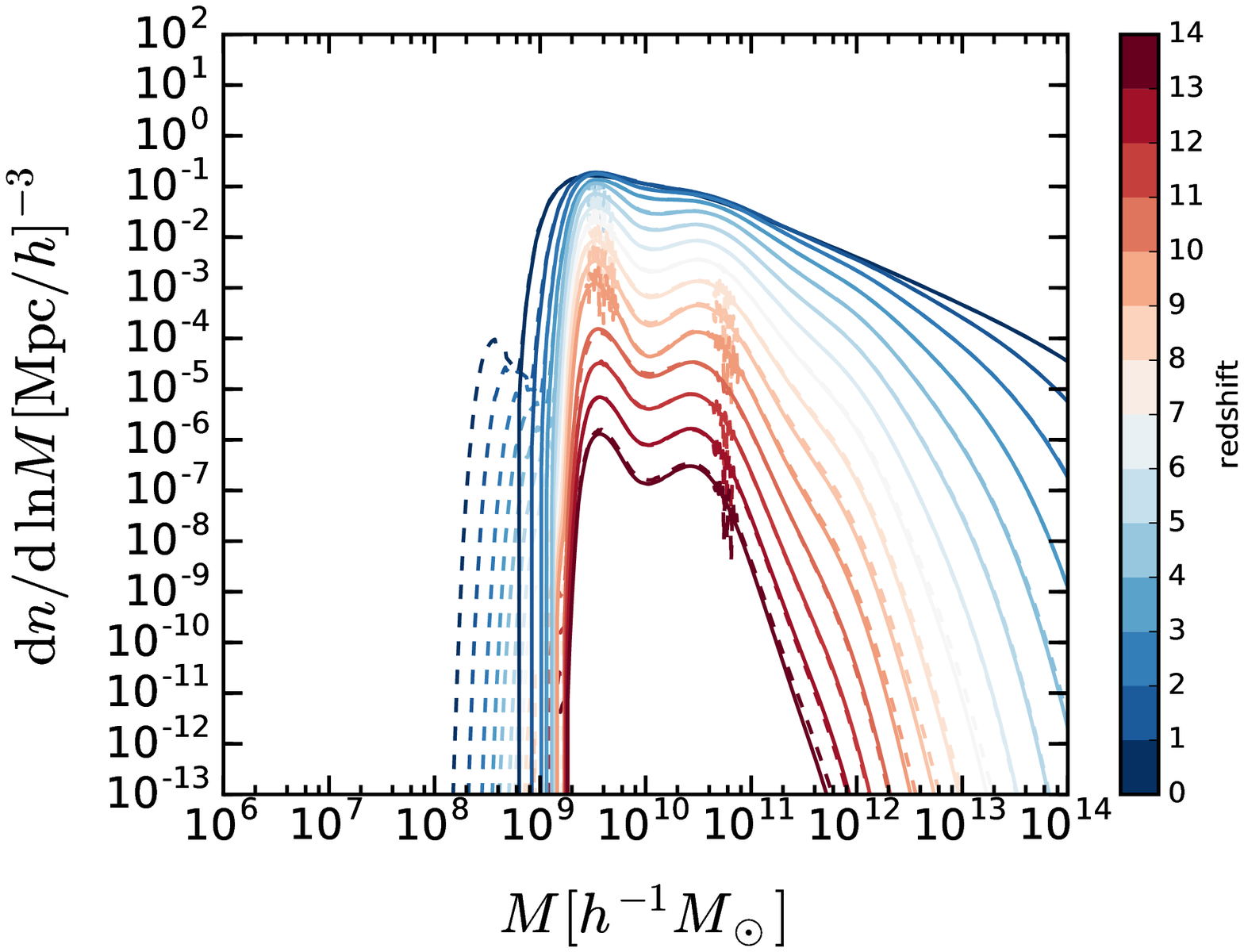}
\caption{HMF for FDM fraction $f = 1.0$ derived using the mid-point rule (solid lines) compared to the result using the
trapezoid rule (dashed lines). Redshifts range from $z=14$ to $0$. The FDM mass is set to $m_{\mathrm{a}}=10^{-22}{\rm eV}$.} 
\label{fig:HMF_com}
\end{figure}

Fig. \ref{fig:HMF_com} shows that the results from the trapezoid rule display artefacts near the cutoff at lower masses. Additionally, we see unphysical oscillations at intermediate masses, which is a common behaviour when one solves
integral equations using the trapezoidal rule \citep{Jones1961,Linz01011969}. Our method avoids these artefacts.

\bsp	
\label{lastpage}

\begin{thebibliography}{}
\makeatletter
\relax
\def\mn@urlcharsother{\let\do\@makeother \do\$\do\&\do\#\do\^\do\_\do\%\do\~}
\def\mn@doi{\begingroup\mn@urlcharsother \@ifnextchar [ {\mn@doi@}
  {\mn@doi@[]}}
\def\mn@doi@[#1]#2{\def\@tempa{#1}\ifx\@tempa\@empty \href
  {http://dx.doi.org/#2} {doi:#2}\else \href {http://dx.doi.org/#2} {#1}\fi
  \endgroup}
\def\mn@eprint#1#2{\mn@eprint@#1:#2::\@nil}
\def\mn@eprint@arXiv#1{\href {http://arxiv.org/abs/#1} {{\tt arXiv:#1}}}
\def\mn@eprint@dblp#1{\href {http://dblp.uni-trier.de/rec/bibtex/#1.xml}
  {dblp:#1}}
\def\mn@eprint@#1:#2:#3:#4\@nil{\def\@tempa {#1}\def\@tempb {#2}\def\@tempc
  {#3}\ifx \@tempc \@empty \let \@tempc \@tempb \let \@tempb \@tempa \fi \ifx
  \@tempb \@empty \def\@tempb {arXiv}\fi \@ifundefined
  {mn@eprint@\@tempb}{\@tempb:\@tempc}{\expandafter \expandafter \csname
  mn@eprint@\@tempb\endcsname \expandafter{\@tempc}}}

\bibitem[\protect\citeauthoryear{Abazajian}{Abazajian}{2006}]{Abazajian:2005xn}
Abazajian K.,  2006, \mn@doi [Phys. Rev. D] {10.1103/PhysRevD.73.063513}, 73,
  063513

\bibitem[\protect\citeauthoryear{{Abbott} et~al.,}{{Abbott}
  et~al.}{2016}]{DES2016}
{Abbott} T.,  et~al., 2016, \mn@doi [\mnras] {10.1093/mnras/stw641}, 460, 1270

\bibitem[\protect\citeauthoryear{Ade et~al.}{Ade et~al.}{2016}]{Planck:2015xua}
Ade P. A.~R.,  et~al., 2016, \mn@doi [\aap] {10.1051/0004-6361/201525830}, 594,
  A13

\bibitem[\protect\citeauthoryear{Arvanitaki, Dimopoulos, Dubovsky, Kaloper  \&
  March-Russell}{Arvanitaki et~al.}{2010}]{Arvanitaki:2009fg}
Arvanitaki A.,  Dimopoulos S.,  Dubovsky S.,  Kaloper N.,   March-Russell J.,
  2010, \mn@doi [Phys. Rev. D] {10.1103/PhysRevD.81.123530}, 81, 123530

\bibitem[\protect\citeauthoryear{{Benson}}{{Benson}}{2005}]{Benson2005}
{Benson} A.~J.,  2005, \mn@doi [\mnras] {10.1111/j.1365-2966.2005.08788.x},
  358, 551

\bibitem[\protect\citeauthoryear{{Benson}}{{Benson}}{2012}]{Benson2012}
{Benson} A.~J.,  2012, \mn@doi [New Astron.] {10.1016/j.newast.2011.07.004},
  17, 175

\bibitem[\protect\citeauthoryear{{Benson} et~al.,}{{Benson}
  et~al.}{2013}]{Benson:2012su}
{Benson} A.~J.,  et~al., 2013, \mn@doi [\mnras] {10.1093/mnras/sts159}, 428,
  1774

\bibitem[\protect\citeauthoryear{Bode, Ostriker  \& Turok}{Bode
  et~al.}{2001}]{Bode:2000gq}
Bode P.,  Ostriker J.~P.,   Turok N.,  2001, \mn@doi [\apj] {10.1086/321541},
  556, 93

\bibitem[\protect\citeauthoryear{Bond, Cole, Efstathiou  \& Kaiser}{Bond
  et~al.}{1991}]{Bond:1990iw}
Bond J.~R.,  Cole S.,  Efstathiou G.,   Kaiser N.,  1991, \mn@doi [\apj]
  {10.1086/170520}, 379, 440

\bibitem[\protect\citeauthoryear{Boylan-Kolchin, Bullock  \&
  Kaplinghat}{Boylan-Kolchin et~al.}{2011}]{Boylan-Kolchin01072011}
Boylan-Kolchin M.,  Bullock J.~S.,   Kaplinghat M.,  2011, \mn@doi [\mnras]
  {10.1111/j.1745-3933.2011.01074.x}, 415, L40

\bibitem[\protect\citeauthoryear{Bozek, Marsh, Silk  \& Wyse}{Bozek
  et~al.}{2015}]{Bozek2015}
Bozek B.,  Marsh D. J.~E.,  Silk J.,   Wyse R. F.~G.,  2015, \mn@doi [\mnras]
  {10.1093/mnras/stv624}, 450, 209

\bibitem[\protect\citeauthoryear{Calabrese \& Spergel}{Calabrese \&
  Spergel}{2016}]{Calabrese2016}
Calabrese E.,  Spergel D.~N.,  2016, \mn@doi [\mnras] {10.1093/mnras/stw1256},
  460, 4397

\bibitem[\protect\citeauthoryear{Cen}{Cen}{2016}]{Cen:2016htm}
Cen R.,  2016, preprint (\mn@eprint {arXiv} {1606.05930})

\bibitem[\protect\citeauthoryear{{Chandrasekhar}}{{Chandrasekhar}}{1943}]{Chandrasekhar1943}
{Chandrasekhar} S.,  1943, \mn@doi [\apj] {10.1086/144517}, 97, 255

\bibitem[\protect\citeauthoryear{Cole, Lacey, Baugh  \& Frenk}{Cole
  et~al.}{2000}]{Cole:2000ex}
Cole S.,  Lacey C.~G.,  Baugh C.~M.,   Frenk C.~S.,  2000, \mn@doi [\mnras]
  {10.1046/j.1365-8711.2000.03879.x}, 319, 168

\bibitem[\protect\citeauthoryear{Gao, Navarro, Cole, Frenk, White, Springel,
  Jenkins  \& Neto}{Gao et~al.}{2008}]{Gao:2007gh}
Gao L.,  Navarro J.~F.,  Cole S.,  Frenk C.,  White S. D.~M.,  Springel V.,
  Jenkins A.,   Neto A.~F.,  2008, \mn@doi [\mnras]
  {10.1111/j.1365-2966.2008.13277.x}, 387, 536

\bibitem[\protect\citeauthoryear{{Gnedin}, {Hernquist}  \& {Ostriker}}{{Gnedin}
  et~al.}{1999}]{Gnedin1999}
{Gnedin} O.~Y.,  {Hernquist} L.,   {Ostriker} J.~P.,  1999, \mn@doi [\apj]
  {10.1086/306910}, 514, 109

\bibitem[\protect\citeauthoryear{{Hezaveh}, {Dalal}, {Holder}, {Kisner},
  {Kuhlen}  \& {Perreault Levasseur}}{{Hezaveh} et~al.}{2014}]{Hezaveh2014}
{Hezaveh} Y.,  {Dalal} N.,  {Holder} G.,  {Kisner} T.,  {Kuhlen} M.,
  {Perreault Levasseur} L.,  2014, preprint (\mn@eprint {arXiv} {1403.2720})

\bibitem[\protect\citeauthoryear{{Hezaveh} et~al.,}{{Hezaveh}
  et~al.}{2016}]{Hezaveh2016}
{Hezaveh} Y.~D.,  et~al., 2016, \mn@doi [\apj] {10.3847/0004-637X/823/1/37},
  823, 37

\bibitem[\protect\citeauthoryear{Hlozek, Grin, Marsh  \& Ferreira}{Hlozek
  et~al.}{2015}]{Hlozek:2014lca}
Hlozek R.,  Grin D.,  Marsh D. J.~E.,   Ferreira P.~G.,  2015, \mn@doi [Phys.
  Rev. D] {10.1103/PhysRevD.91.103512}, 91, 103512

\bibitem[\protect\citeauthoryear{Hu, Barkana  \& Gruzinov}{Hu
  et~al.}{2000}]{Hu:2000ke}
Hu W.,  Barkana R.,   Gruzinov A.,  2000, \mn@doi [Phys. Rev. Lett.]
  {10.1103/PhysRevLett.85.1158}, 85, 1158

\bibitem[\protect\citeauthoryear{{Jiang}, {Jing}, {Faltenbacher}, {Lin}  \&
  {Li}}{{Jiang} et~al.}{2008}]{Jiang2008}
{Jiang} C.~Y.,  {Jing} Y.~P.,  {Faltenbacher} A.,  {Lin} W.~P.,   {Li} C.,
  2008, \mn@doi [\apj] {10.1086/526412}, 675, 1095

\bibitem[\protect\citeauthoryear{Jones}{Jones}{1961}]{Jones1961}
Jones J.~G.,  1961, Math. Comput., 15, 131

\bibitem[\protect\citeauthoryear{Khlopov, Malomed  \& Zeldovich}{Khlopov
  et~al.}{1985}]{Khlopov01081985}
Khlopov M.~Y.,  Malomed B.~A.,   Zeldovich Y.~B.,  1985, \mn@doi [\mnras]
  {10.1093/mnras/215.4.575}, 215, 575

\bibitem[\protect\citeauthoryear{{King}}{{King}}{1962}]{King1962}
{King} I.,  1962, \mn@doi [\apj] {10.1086/108756}, 67, 471

\bibitem[\protect\citeauthoryear{Klypin, Kravtsov, Valenzuela  \& Prada}{Klypin
  et~al.}{1999}]{Klypin:1999uc}
Klypin A.~A.,  Kravtsov A.~V.,  Valenzuela O.,   Prada F.,  1999, \mn@doi
  [\apj] {10.1086/307643}, 522, 82

\bibitem[\protect\citeauthoryear{Linz}{Linz}{1969}]{Linz01011969}
Linz P.,  1969, \mn@doi [Comput. J.] {10.1093/comjnl/12.4.393}, 12, 393

\bibitem[\protect\citeauthoryear{Lora \& Maga{\~n}a}{Lora \&
  Maga{\~n}a}{2014}]{Lora:2014cya}
Lora V.,  Maga{\~n}a J.,  2014, \mn@doi [\jcap]
  {10.1088/1475-7516/2014/09/011}, 1409, 011

\bibitem[\protect\citeauthoryear{Lora, Magana, Bernal, Sanchez-Salcedo  \&
  Grebel}{Lora et~al.}{2012}]{Lora:2011yc}
Lora V.,  Magana J.,  Bernal A.,  Sanchez-Salcedo F.~J.,   Grebel E.~K.,  2012,
  \mn@doi [\jcap] {10.1088/1475-7516/2012/02/011}, 1202, 011

\bibitem[\protect\citeauthoryear{Marsh}{Marsh}{2016a}]{Marsh2016}
Marsh D. J.~E.,  2016a, personal communication

\bibitem[\protect\citeauthoryear{Marsh}{Marsh}{2016b}]{Marsh:2016vgj}
Marsh D. J.~E.,  2016b, preprint (\mn@eprint {arXiv} {1605.05973})

\bibitem[\protect\citeauthoryear{Marsh}{Marsh}{2016c}]{Marsh2015}
Marsh D. J.~E.,  2016c, \mn@doi [Phys. Rept.] {10.1016/j.physrep.2016.06.005},
  643, 1

\bibitem[\protect\citeauthoryear{{Marsh} \& {Pop}}{{Marsh} \&
  {Pop}}{2015}]{Marsh2015b}
{Marsh} D.~J.~E.,  {Pop} A.-R.,  2015, \mn@doi [\mnras]
  {10.1093/mnras/stv1050}, 451, 2479

\bibitem[\protect\citeauthoryear{Marsh \& Silk}{Marsh \&
  Silk}{2014}]{Marsh2014}
Marsh D. J.~E.,  Silk J.,  2014, \mn@doi [\mnras] {10.1093/mnras/stt2079}, 437,
  2652

\bibitem[\protect\citeauthoryear{{Monaco}}{{Monaco}}{1998}]{Monaco1998}
{Monaco} P.,  1998, Fundam. Cosm. Phys., 19, 157

\bibitem[\protect\citeauthoryear{Moore, Ghigna, Governato, Lake, Quinn, Stadel
  \& Tozzi}{Moore et~al.}{1999}]{Moore:1999nt}
Moore B.,  Ghigna S.,  Governato F.,  Lake G.,  Quinn T.~R.,  Stadel J.,
  Tozzi P.,  1999, \mn@doi [\apj] {10.1086/312287}, 524, L19

\bibitem[\protect\citeauthoryear{Navarro, Frenk  \& White}{Navarro
  et~al.}{1996}]{Navarro:1995iw}
Navarro J.~F.,  Frenk C.~S.,   White S. D.~M.,  1996, \mn@doi [\apj]
  {10.1086/177173}, 462, 563

\bibitem[\protect\citeauthoryear{Navarro, Frenk  \& White}{Navarro
  et~al.}{1997}]{Navarro:1996gj}
Navarro J.~F.,  Frenk C.~S.,   White S.~D.,  1997, \mn@doi [\apj]
  {10.1086/304888}, 490, 493

\bibitem[\protect\citeauthoryear{Parkinson, Cole  \& Helly}{Parkinson
  et~al.}{2008}]{Parkinson:2007yh}
Parkinson H.,  Cole S.,   Helly J.,  2008, \mn@doi [\mnras]
  {10.1111/j.1365-2966.2007.12517.x}, 383, 557

\bibitem[\protect\citeauthoryear{Press \& Schechter}{Press \&
  Schechter}{1974}]{Press:1973iz}
Press W.~H.,  Schechter P.,  1974, \mn@doi [\apj] {10.1086/152650}, 187, 425

\bibitem[\protect\citeauthoryear{Pullen, Benson  \& Moustakas}{Pullen
  et~al.}{2014}]{Pullen:2014gna}
Pullen A.~R.,  Benson A.~J.,   Moustakas L.~A.,  2014, \mn@doi [\apj]
  {10.1088/0004-637X/792/1/24}, 792, 24

\bibitem[\protect\citeauthoryear{Sahni \& Wang}{Sahni \&
  Wang}{2000}]{Sahni:1999qe}
Sahni V.,  Wang L.-M.,  2000, \mn@doi [Phys. Rev. D]
  {10.1103/PhysRevD.62.103517}, 62, 103517

\bibitem[\protect\citeauthoryear{{Sanders}, {Bovy}  \& {Erkal}}{{Sanders}
  et~al.}{2016}]{Sanders2016}
{Sanders} J.~L.,  {Bovy} J.,   {Erkal} D.,  2016, \mn@doi [\mnras]
  {10.1093/mnras/stw232}, 457, 3817

\bibitem[\protect\citeauthoryear{{Sarkar}, {Mondal}, {Das}, {Sethi},
  {Bharadwaj}  \& {Marsh}}{{Sarkar} et~al.}{2016}]{Sarkar2016}
{Sarkar} A.,  {Mondal} R.,  {Das} S.,  {Sethi} S.~K.,  {Bharadwaj} S.,
  {Marsh} D.~J.~E.,  2016, \mn@doi [\jcap] {10.1088/1475-7516/2016/04/012}, 4,
  012

\bibitem[\protect\citeauthoryear{{Schive}, {Chiueh}  \& {Broadhurst}}{{Schive}
  et~al.}{2014a}]{Schive2014a}
{Schive} H.-Y.,  {Chiueh} T.,   {Broadhurst} T.,  2014a, \mn@doi [Nat. Phys.]
  {10.1038/nphys2996}, 10, 496

\bibitem[\protect\citeauthoryear{{Schive}, {Liao}, {Woo}, {Wong}, {Chiueh},
  {Broadhurst}  \& {Hwang}}{{Schive} et~al.}{2014b}]{Schive2014b}
{Schive} H.-Y.,  {Liao} M.-H.,  {Woo} T.-P.,  {Wong} S.-K.,  {Chiueh} T.,
  {Broadhurst} T.,   {Hwang} W.-Y.~P.,  2014b, \mn@doi [Phys. Rev. Lett.]
  {10.1103/PhysRevLett.113.261302}, 113, 261302

\bibitem[\protect\citeauthoryear{{Schive}, {Chiueh}, {Broadhurst}  \&
  {Huang}}{{Schive} et~al.}{2016}]{Schive2016}
{Schive} H.-Y.,  {Chiueh} T.,  {Broadhurst} T.,   {Huang} K.-W.,  2016, \mn@doi
  [\apj] {10.3847/0004-637X/818/1/89}, 818, 89

\bibitem[\protect\citeauthoryear{Schneider}{Schneider}{2015}]{Schneider:2014rda}
Schneider A.,  2015, \mn@doi [\mnras] {10.1093/mnras/stv1169}, 451, 3117

\bibitem[\protect\citeauthoryear{Schneider, Smith, Maccio  \& Moore}{Schneider
  et~al.}{2012}]{Schneider:2011yu}
Schneider A.,  Smith R.~E.,  Maccio A.~V.,   Moore B.,  2012, \mn@doi [\mnras]
  {10.1111/j.1365-2966.2012.21252.x}, 424, 684

\bibitem[\protect\citeauthoryear{Schneider, Smith  \& Reed}{Schneider
  et~al.}{2013}]{Schneider:2013ria}
Schneider A.,  Smith R.~E.,   Reed D.,  2013, \mn@doi [\mnras]
  {10.1093/mnras/stt829}, 433, 1573

\bibitem[\protect\citeauthoryear{Schwabe, Niemeyer  \& Engels}{Schwabe
  et~al.}{2016}]{Schwabe2016}
Schwabe B.,  Niemeyer J.~C.,   Engels J.~F.,  2016, \mn@doi [Phys. Rev. D]
  {10.1103/PhysRevD.94.043513}, 94, 043513

\bibitem[\protect\citeauthoryear{Sheth \& Tormen}{Sheth \&
  Tormen}{1999}]{Sheth:1999mn}
Sheth R.~K.,  Tormen G.,  1999, \mn@doi [\mnras]
  {10.1046/j.1365-8711.1999.02692.x}, 308, 119

\bibitem[\protect\citeauthoryear{Sheth, Mo  \& Tormen}{Sheth
  et~al.}{2001}]{Sheth:1999su}
Sheth R.~K.,  Mo H.~J.,   Tormen G.,  2001, \mn@doi [\mnras]
  {10.1046/j.1365-8711.2001.04006.x}, 323, 1

\bibitem[\protect\citeauthoryear{Spergel \& Steinhardt}{Spergel \&
  Steinhardt}{2000}]{Spergel:1999mh}
Spergel D.~N.,  Steinhardt P.~J.,  2000, \mn@doi [Phys. Rev. Lett.]
  {10.1103/PhysRevLett.84.3760}, 84, 3760

\bibitem[\protect\citeauthoryear{Svrcek \& Witten}{Svrcek \&
  Witten}{2006}]{Svrcek:2006yi}
Svrcek P.,  Witten E.,  2006, \mn@doi [J. High Energy Phys.]
  {10.1088/1126-6708/2006/06/051}, 06, 051

\bibitem[\protect\citeauthoryear{{Walker} \& {Pe{\~n}arrubia}}{{Walker} \&
  {Pe{\~n}arrubia}}{2011}]{Walker2011}
{Walker} M.~G.,  {Pe{\~n}arrubia} J.,  2011, \mn@doi [\apj]
  {10.1088/0004-637X/742/1/20}, 742, 20

\bibitem[\protect\citeauthoryear{Wetzel, Hopkins, Kim, Faucher-Giguere, Keres
  \& Quataert}{Wetzel et~al.}{2016}]{Wetzel:2016wro}
Wetzel A.~R.,  Hopkins P.~F.,  Kim J.-h.,  Faucher-Giguere C.-A.,  Keres D.,
  Quataert E.,  2016, \mn@doi [\apj] {10.3847/2041-8205/827/2/L23}, 827, L23

\bibitem[\protect\citeauthoryear{Witten}{Witten}{1984}]{Witten:1984dg}
Witten E.,  1984, \mn@doi [Phys. Lett. B] {10.1016/0370-2693(84)90422-2}, 149,
  351

\bibitem[\protect\citeauthoryear{Woo \& Chiueh}{Woo \&
  Chiueh}{2009}]{Woo:2008nn}
Woo T.-P.,  Chiueh T.,  2009, \mn@doi [\apj] {10.1088/0004-637X/697/1/850},
  697, 850

\bibitem[\protect\citeauthoryear{de Blok}{de~Blok}{2010}]{deBlok:2009sp}
de Blok W.,  2010, \mn@doi [Adv. Astron.] {10.1155/2010/789293}, 2010, 789293

\makeatother
\end{thebibliography}
\end{document}